\documentclass[useAMS,usenatbib,usegraphicx]{mn2e}
\usepackage{aasmacros,url,units}
\usepackage{amssymb,amsfonts,amsmath,paralist,xspace} %

\title[Draco DM X-ray signal]{Decaying dark matter: the case for a deep X-ray observation of Draco}
\author[Mark R. Lovell~et.~al.]{Mark R. Lovell$^{1,2}$\thanks{E-mail:M.R.Lovell@uva.nl},
Gianfranco Bertone$^{1}$, Alexey Boyarsky$^{2}$, Adrian Jenkins$^{3}$,\newauthor
 and Oleg Ruchayskiy$^{4}$\\
$^{1}$GRAPPA, Universiteit van Amsterdam, Science Park 904, 1098 XH
Amsterdam, The Netherlands\\
$^{2}$Instituut-Lorentz for Theoretical Physics, Niels Bohrweg 2, Leiden, The Netherlands\\
$^{3}$Institute for Computational Cosmology, Durham University, South
Road, Durham, UK, DH1 3LE\\
$^{4}$\'Ecole Polytechnique F\'ed\'erale de Lausanne, FSB/ITP/LPPC, BSP 720, CH-1015,
Lausanne, Switzerland\\}

\newcommand{\Msun}{\mathrm{M_{\sun}}}

\newcommand{\dm}{{\scriptscriptstyle\rm DM}}

\def\gsim{ \lower .75ex \hbox{$\sim$} \llap{\raise .27ex \hbox{$>$}} }
\def\lsim{ \lower .75ex \hbox{$\sim$} \llap{\raise .27ex \hbox{$<$}} }

\def\jcap{JCAP}
\begin{document}

  \date{Accepted *** Received ***; in original
    form ***} 

  \pagerange{\pageref{firstpage}--\pageref{lastpage}} \pubyear{2014}

  \maketitle

  \label{first page}

  \begin{abstract}
    Recent studies of M31, the Galactic centre, and galaxy clusters have made
    tentative detections of an X-ray line at $\sim 3.5$~keV that could be
    produced by decaying dark matter. We use high resolution simulations of
    the Aquarius project to predict the likely amplitude of the X-ray decay
    flux observed in the GC relative to that observed in M31, and also of the
    GC relative to other parts of the Milky Way halo and to dwarf spheroidal
    galaxies. We show that the reported detections from M31 and the GC are
    compatible with each other, and with upper limits arising from high
    galactic latitude observations, and imply a decay time $\tau \sim 10^{28}$
    seconds. We argue that this interpretation can be tested with deep
    observations of dwarf spheroidal galaxies: in 95 per cent of our mock
    observations, a 1.3 Msec pointed observation of Draco with XMM-Newton will
    enable us to discover or rule out at the 3$\sigma$ level an X-ray feature
    from dark matter decay at 3.5 keV, for decay times $\tau < 0.8 \times
    10^{28}$~sec.
    
  \end{abstract}


  \section{Introduction}
  \label{intro}

  One of the most pressing and interesting questions in fundamental physics
  and cosmology today is the identity of the dark matter (see
  e.g. \citealt{Bertone:2010zza} and refs.\ therein).  The properties of
  hypothetical dark matter particles remain largely unknown and there are many
  attempts to find phenomenological constraints on particular models from
  astronomical observations.  For example, if dark matter particles were born
  relativistic, deep in the radiation-dominated epoch (so called ``warm dark
  matter'', or WDM) they would affect the way structures were formed at small
  scales, leaving their imprints in the Lyman-$\alpha$ forest
  \citep[e.g.][]{Viel05,Boyarsky2009c,Viel_13} and satellite galaxy abundances
  and structure \citep{Polisensky2011,Kennedy14,Lovell14}.  Satellite
  structure may also provide limits on dark matter self-interactions
  \citep{Vogelsberger12,Zavala13}.  One further constraint is the detection of
  electromagnetic radiation originating from the decay or annihilation of dark
  matter particles. Many studies to date have centred on attempts to detect
  the annihilation of weakly interacting massive particles (WIMPs)~\citep[see
  e.g.][for a review]{Bertone:2004pz,Feng:2010gw}. This generic class of particles
  is attractive as a dark matter candidate due to their stability, their
  potential relation to electroweak symmetry breaking, and the possibility
  that they may be detected in laboratory experiments \citep[see e.g.][and
  refs.\ therein]{Cerdeno2010}. WIMPs cannot decay, but are predicted to
  annihilate with one another in regions of high dark matter density, and
  could be detected via their annihilation products. 
  Currently, the most interesting candidate WIMP signal is
  an excess of GeV photons from the centre of the Galaxy
  \citep{Hooper11,Daylan:2014rsa,Calore:2014xka}, but further evidence is
  needed to rule out a possible astrophysical origin of the signal
  \citep{Boyarsky11}.

  The interaction strength of weakly interacting massive particles limits
  their mass to the few GeV--few TeV range (in order to give the correct
  primordial abundance). Once the assumption about the interaction strength is
  relaxed, particle physics theories predict the existence of dark matter of
  different masses. If the particle mass is at the keV scale, and its lifetime
  longer than the age of the Universe, it may in principle be detectable
  indirectly, through the observation of X-ray photons produced by its
  decay~\citep[see e.g.\
  ][]{Dolgov:00,Abazajian:01b,Boyarsky:06c,denHerder:09,Abazajian:09a}. An
  X-ray line signal consistent with such a decay has been identified at an
  energy of $\sim 3.5$~keV in galaxy clusters, in the Milky Way centre (or
  Galactic centre, GC) and in M31 by several studies
  \citep{Boyarsky14,Boyarsky14b,Bulbul14}.\footnote{Both the Milky
    Way~\citep{Abazajian:06b,Boyarsky:06d,Riemer:06} and
    M31~\citep{Watson:06,Boyarsky:07a,Boyarsky:10b,Watson:11,Horiuchi:13} have
    been extensively studied in this aspect. However, each of the datasets
    used in previous decaying DM searches has poorer statistics than used in
    \cite{Boyarsky14,Boyarsky14b,Jeltema14,RiemerSorensen14}. The
    non-detection of any signal in these works does not contradict
    current results.} although others have reported non-detections
  \citep{Anderson14, Malyshev14} or reported detections but attributed them to
  have astrophysical origins (\citealp{Jeltema14,RiemerSorensen14} but see
  also \citealp{Boyarsky14c, Bulbul14b}). These two classes of systems --
  clusters and $L_{*}$ galaxies -- are separated in mass by three orders of
  magnitude, therefore the correlation of the measured signal with expected
  projected dark matter mass is compelling evidence for dark matter decay as
  the origin of the signal.

  These studies have based their estimates of the dark matter
  distribution in their targets on dark halo mass models that do not take into
  account fully the triaxiality of the halo, the presence of substructure,
  or the effects of baryons. Some of these issues have been examined
  in \citet{Bernal14}, who used a low resolution cosmological
  simulation containing
  $\sim10^{5}$ Milky Way halo analogues to motivate better triaxial
  halo mass models. We instead make use of a series of high
  resolution simulations of Milky Way-analogue dark matter haloes, some
  of which were run with a full hydrodynamical treatment of the
  baryonic component, to estimate the X-ray decay signal from
  these targets and compare the results to the reported
  detections. 

  Targets such as the Milky Way and M31 are attractive due to their large
  projected mass densities, however their analysis is complicated by the
  presence of X-ray emission lines from the interstellar medium. Therefore,
  the nature of the line -- dark matter or astrophysical -- may be better
  ascertained by performing observations of objects with much cleaner
  backgrounds. One particularly promising class of candidates is that of the
  Milky Way's dwarf spheroidal satellites~\citep{Boyarsky:06c}. These galaxies
  have very high mass-to-light ratios \citep{Walker09, Walker10, Wolf10}, and
  very low gas fractions \citep[][and references therein]{Gallagher03}. Their
  X-ray emitting gas fractions will be lower still due to their small gas
  fractions. Any detection from these galaxies would thus have a very low
  probability of an astrophysical origin and therefore dwarf spheroidal
  satellites have previously been targets of decaying DM searches in X-rays
  \citep{Boyarsky:06c,Boyarsky:06d,Loewenstein:08,Riemer:09a,Loewenstein:09,Mirabal:10b,Loewenstein:12,Malyshev14}.
  With
  our set of simulations we can also set out to calculate the flux from dwarf
  spheroidals in their full cosmological context, including the contribution
  from the host halo.
  
  In this work we use the high resolution simulations of the Aquarius
  project \citep{Springel08b}
  to predict the likely amplitude of the X-ray decay flux observed in the GC
  relative to that observed in M31, and also of the GC relative to other parts
  of the Milky Way halo and to dwarf spheroidal galaxies.  As we shall see, a
  sufficiently deep observation of the Draco dwarf galaxy would allow us to
  test the dark matter interpretation of the observed X-ray line signal in a
  very large region of the parameter space.

  This paper is
  organised as follows. In Section~\ref{sims} we present the
  simulations used in this paper. We discuss our methods for
  calculating the X-ray flux from the simulations in
  Section~\ref{MXS}. In Section~\ref{Res} we present our results for
  the expected fluxes of the GC, M31, and two dwarf spheroidal
  galaxies,  and draw conclusions in
  Section~\ref{Conc}. In the appendices we test the convergence of the
  simulations as function of resolution and number of sightlines
  (Appendix~\ref{RT}), and consider the effects of cosmology,
  baryonic physics, and dark matter power spectrum
  (Appendix~\ref{MI}).  We present observational analysis and
  predictions for signal from Draco in Appendix~\ref{DO}. 

  \section{Simulations}
  \label{sims}

  \begin{table*}
    \centering
    \begin{tabular}{|l|c|c|c|c|c|}
      \hline
      Simulation & $m_{p}$ [$\Msun$] & $\epsilon$ [pc] & $M_{200}$
      [$\Msun$]& $c$ & $m_{\rmn{WDM}} [\rmn{keV}]$ \\
      \hline  
      Aq-A1 & $1.712\times10^{3}$ & 20.5  & $1.839\times10^{12}$ &
      18.6 & -- \\
      Aq-A2 & $1.370\times10^{4}$ & 65.8  & $1.842\times10^{12}$ &
      18.5 & -- \\
      Aq-A3 & $4.911\times10^{4}$ & 120.5 & $1.836\times10^{12}$ &
      18.5 & -- \\
      Aq-A4 & $3.929\times10^{5}$ & 342.5 & $1.838\times10^{12}$ &
      18.6 & -- \\[0.25cm]
      Aq-A2(W7) & $1.545\times10^{4}$ & 68.2 & $1.938\times10^{12}$ &
      16.1 & -- \\
      Aq-A2--$m_{1.5}$(W7) & $1.545\times10^{4}$ & 68.2  &
      $1.797\times10^{12}$ & 15.9 & 1.456 \\
      Aq-A2--$m_{1.6}$(W7) & $1.545\times10^{4}$ & 68.2 &
      $1.802\times10^{12}$ & 16.2 & 1.637 \\
      Aq-A2--$m_{2.0}$(W7) & $1.545\times10^{4}$ & 68.2  &
      $1.843\times10^{12}$ & 16.0 & 2.001 \\
      Aq-A2--$m_{2.3}$(W7) & $1.545\times10^{4}$ & 68.2 &
      $1.875\times10^{12}$ & 16.1 & 2.322 \\[0.25cm]
      Aq-A4-S-NoWinds & $3.222\times10^{5}$ & 342.5 &
      $1.709\times10^{12}$ & 25.1 & -- \\
      Aq-A4-S-Winds & $3.222\times10^{5}$ & 342.5 &
      $1.590\times10^{12}$ & 27.5 & -- \\[0.25cm]
      Aq-B2 & $6.447\times10^{3}$ & 65.8  & $8.194\times10^{11}$ & 11.7
      & -- \\
      Aq-C2 & $1.399\times10^{4}$ & 65.8  & $1.774\times10^{12}$ & 18.4
      & -- \\
      Aq-D2 & $1.397\times10^{4}$ & 65.8  & $1.774\times10^{12}$ & 12.4
      & -- \\
      Aq-E2 & $9.593\times10^{3}$ & 65.8  & $1.185\times10^{12}$ & 15.3
      & -- \\

      \hline
    \end{tabular}
    \caption{Parameters of the 
      simulations. We include the simulation dark matter particle mass
      $m_{p}$, the smoothing length $\epsilon$, the mass of the central halo encompassing a
      region of 200 times the critical density of the Universe,
      $M_{200}$, the halo concentration $c$, and the WDM particle mass,
      $m_{\rmn{WDM}}$, where applicable. Concentration is determined
      by fitting NFW profiles to each halo at radii between 1~kpc and 100~kpc.}
    \label{TabSim}
  \end{table*}

  Each of the simulations used in this study is either taken directly from or derived from the
  Aquarius project. This is a set of Milky Way
  halo-analogue dark matter-only simulations run with the {\sc
  p-gadget3} code \citep{Springel08b}, and uses
  the cosmological parameters consistent with the one-year data of the
  {\it Wilkinson Microwave
  Anisotropy Probe (WMAP1)}: $H_{0}=100h=73\rmn{km/s/Mpc}$,
  $\Omega_{\rmn{m}}=0.25$, $\Omega_{\Lambda}=0.75$, $\sigma_{8}=0.9$, and
  $n_{s}=1$ \citep{wmap1}. The six haloes are labelled Aq-A through to
  Aq-F. Aq-F was found to experience a major merger at redshift
  0.4-0.5, and has been shown to be likely to host an S0
  galaxy rather than a disc galaxy at redshift
  zero \citep{Cooper11}: we therefore do not consider it in this
  study. The remaining five haloes span a range of masses and
  concentrations, and thus enable us to examine the effect of these
  parameters for different possible values of Milky Way and M31 mass
  and concentration. 

  The Aq-A halo has been rerun at five different
  simulation resolution
  levels for the purpose of checking that results are converged
  and not influenced by simulation resolution. These levels are
  labelled from 1 (highest resolution, particle
  mass $\sim10^{3}\Msun$) to 5 (lowest resolution, particle
  mass $\sim3\times10^{6}\Msun$). The precise particle masses are
  reproduced in Table~\ref{TabSim}. Aq-A5 is very poorly resolved for the
  purposes of this experiment and is therefore not used here.  
  
  We determine the properties of DM structures using the
  \textsc{subfind} algorithm \citep{Springel01}, which identifies bound overdensities as
  haloes and subhaloes. The largest \textsc{subfind} halo in each of the simulations
  is referred to as the `main halo', and we take the Milky Way / M31
  halo centre to be that of the main halo's centre-of-potential. The
  positions of dwarf spheroidal candidates are likewise the
  centres-of-potential of dark matter subhaloes.  

  The original Aquarius runs are dark matter-only simulations, however
  it is likely that baryonic physics will have an impact on the distribution of
  dark matter in the Galaxy. We therefore make use of two gas physics
  resimulations of Aq-A4. Both have been run with the
  \textsc{p-gadget3} code \citep{Springel08b} and adopt the
  gas physics and star-formation prescriptions of
  \citet{Springel03}. The two runs differ only in that one has the
  galactic winds model of \citet{Springel03} enabled whereas the
  other does not. The inclusion of winds inhibits further
  star-formation such that the stellar mass of the central galaxy is
  reduced from $1.45\times10^{11}\Msun$ in the no-winds case to
  $9.19\times10^{10}\Msun$ when winds are included. Both of these
  values are higher than the Milky Way stellar mass inferred by
  \citet[][$6.43\pm0.63\times10^{10}\Msun$]{McMillan11}, so care
  should be taken when comparing these models to the Milky Way,
  especially the model that does not make use of the winds physics.

  To check for the likely effect of our choice of cosmological parameters we have
  also performed a resimulation of Aq-A2 that instead uses the {\it WMAP7} year
  values: $H_{0}=100h=70.4\rmn{km/s/Mpc}$,
  $\Omega_{m}=0.272$, $\Omega_{\Lambda}=0.728$, $\sigma_{8}=0.81$, and
  $n_{s}=0.967$ \citep{wmap11}. Since one candidate for producing
  the decay line is a 7.1keV sterile neutrino, which has the kinematic
  properties of WDM, we make use of four WDM simulations of
  Aq-A2 in the {\it WMAP7} cosmology. These are identical to the
  Aq-A2-{\it WMAP7} run except that the initial conditions wave
  amplitudes are rescaled with thermal relic
  WDM power spectra \citep{Bode01,Viel05}. The WDM models used have
  thermal relic particle masses 1.5keV, 1.6keV, 2.0keV, and
  2.3keV. The 2.0keV model is a good approximation to a sterile
  neutrino produced in the presence of both very low and very high
  lepton asymmetries \citep[][Lovell~et~al.~in~prep.]{Abazajian14},
  and the other three enable us to examine the effect of larger and
  smaller effective particle masses. Further details
  about these simulations and the definition of WDM particle mass may
  be found in \citet{Lovell14}. Important properties for each of the
  simulations are given in Table~\ref{TabSim}.

  \section{Modelling the X-ray signal}
  \label{MXS}

  We treat each simulation dark matter particle as a source of
  X-ray photons, which are emitted uniformly in all directions. It is
  simple to show that, for an inverse decay width $\tau$, the rate of
  photon production via this channel is: 

  \begin{equation}
    \frac{dn}{dt} = \frac{N_{0}}{\tau}~,
  \end{equation}

  \noindent
  where $N_{0}$ is the (initial) number of dark matter particles
  that are represented by each simulation dark matter particle. We will adopt
  $\tau=10^{28}~\rmn{s}$, which is close to the preferred value of
  \citet{Boyarsky14}, for all of the results in this study except
  where indicated otherwise. In order to calculate the number of dark
  matter particles per simulation particle we set the particle mass to
  equal 7.1~keV, since this is the particle mass that would result in
  a two body decay for the 3.55~keV line. We also assume for our dark matter-only
  runs that baryons and dark matter have the same spatial
  distribution, and take the projected dark matter mass to be the
  universal dark matter mass fraction
  $\Omega_{\rmn{DM}}/\Omega_{\rmn{m}}$ multiplied by the total
  projected mass: we apply this fraction to all of our measurements that use dark
  matter-only simulations, including those for dwarf spheroidals.


  We make use of two methods for calculating the flux from our
  chosen targets:  sightlines for particular observer positions,
  and the spherically averaged flux for an observer at a
  given distance from the target. Both of
  these are discussed in detail below.

  \subsection{Sightlines}
  \label{SSsl}
  In this method we treat each simulation particle as a point
  source of X-ray photons. We randomly select positions on the
  surface of a sphere of some radius (8~kpc for the GC, 780~kpc
  for M31) around the centre of the largest \textsc{subfind} halo, and
  place our observers at these positions. We
  then define a cone with an opening angle of the XMM-Newton
  MOS field-of-view (FoV; $14'$ radius) and central axis connecting the observer to the
  halo centre, and calculate the total flux of all simulation particles
  found within that cone: this is our `sightline' measurement. For the
  measurements in which we observe the GC or M31 (`on-centre'
  measurements), the cone is directed towards the halo
  centre as described above; the procedures for `off-centre' and dwarf
  spheroidal observations are described in Section~\ref{Res}.

  All of our runs are zoom simulations, in
  which the halo of interest resides within a region of
  diameter~$\sim2$~Mpc populated with high resolution particles; the
  remainder of the box contains more massive, low resolution particles
  to provide the correct large scale forces. We
  use only the high resolution particles for our study. We find
  that our results are insensitive to any edge effects where the high
  resolution region ends, and we can truncate the region of particles
  sampled down to 50~kpc from the halo centre without affecting any of our
  results: we therefore do not impose any truncation. Background
  dark matter sources beyond the high resolution region will make a
  negligible contribution due to the redshifting of the emission, and
  are thus not included in the analysis \citep{Boyarsky06}. In all
  cases we take the centre of the Milky Way
  (Sagittarius~A*) to be at the simulated halo centre-of-potential as
  determined by our halo finder. We discuss briefly the number of
  sightlines required for our results to be robust in Section~\ref{RT}.

  \subsection{Spherically averaged flux calculation}

  In this method we smear out each simulation particle into a
  spherical shell around the halo centre. We then calculate the flux
  from the shell surface area that intersects the line-of-sight
  cone as described above. This approach is equivalent to using an
  infinite number of sightlines. The numerical noise is reduced relative
  to the sightline method, with the penalty that we wash out
  anisotropies due to halo triaxiality and substructure.   

  It can be shown that the surface area of a spherical shell section
  contained within an observer FoV $\alpha$ situated a distance $d$
  away from the shell centre is approximately:
 
  \begin{equation}
    A \approx 2\pi r(r\pm \alpha^{2}d-\sqrt{r^{2}-\alpha^{2}d^{2}})~,
  \end{equation}
  where $r$ is the sphere radius. The two solutions
  correspond to the surfaces intersected on either side of the shell: if
  the observer is located within the shell then only the solution with
  the plus sign is physical. We treat shells that fit entirely within
  the cone as point sources. The flux from the two portions is then:
  \begin{equation}
      F \approx
      \frac{N_0}{\tau}\frac{r\pm d\alpha^2-\sqrt{r^{2}-d^{2}\alpha^{2}}}{8\pi r(d\pm r)^{2}}~,
  \end{equation}
  and we can sum over all particles in the simulation to obtain the
  spherically-averaged flux of our target. This method is used in the
  appendix only, as a check for our sightline methods and to examine
  how the flux changes with distance to the halo centre.

 \section{Results}
 \label{Res}

 We now consider the effect of comparing different observations:
 on-centre vs. off-centre observations of the GC, on-centre GC
 vs. on-centre M31, and on-centre GC vs two dwarf spheroidal
 galaxies. All observers are placed at randomly selected positions
 around the centre of the target halo as described in
 subsection~\ref{SSsl}.

 \subsection{On-centre vs. off-centre observations}
\begin{figure}
   \includegraphics[scale=0.35,angle=-90]{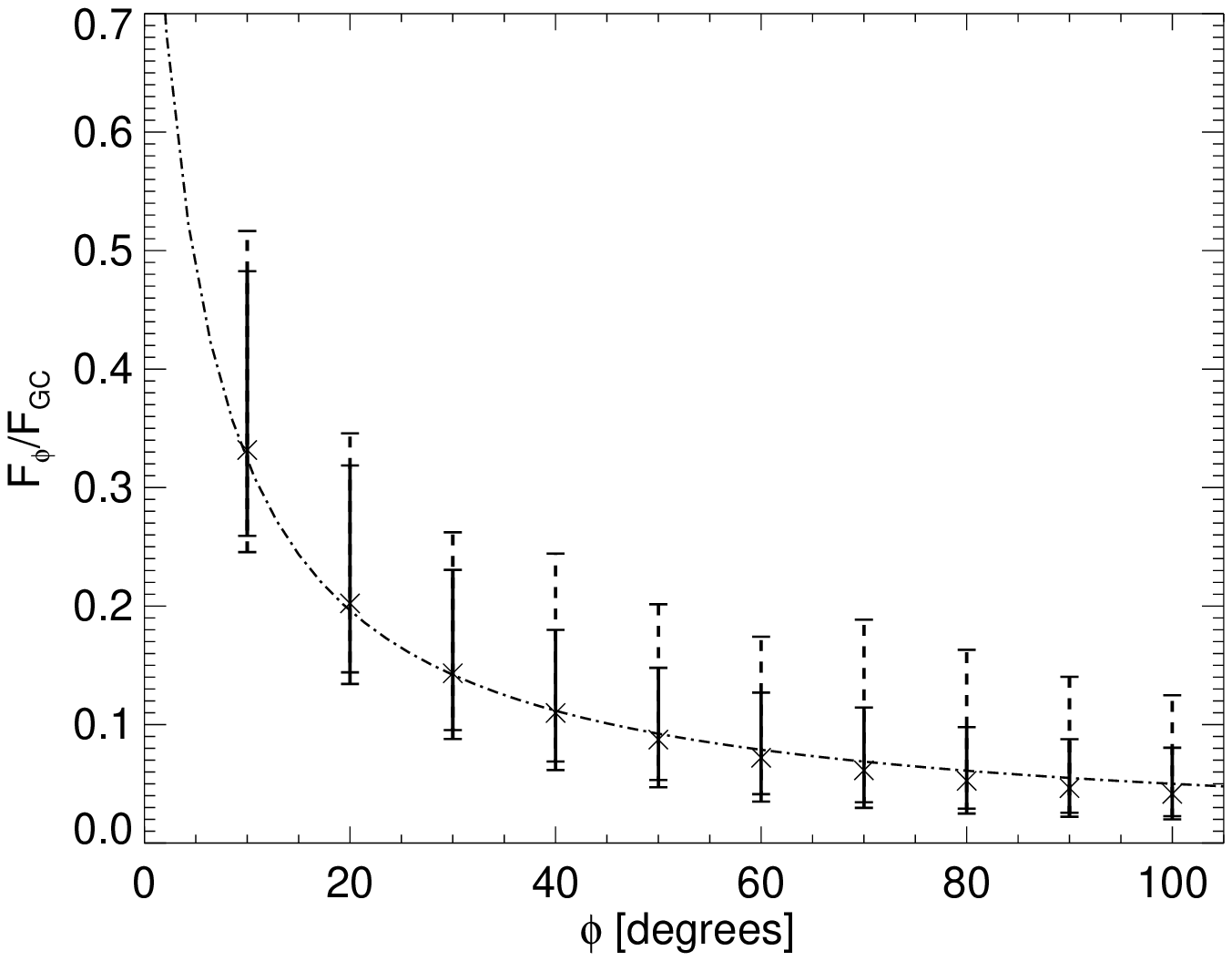}  
   \caption{The ratio of off-centre flux to on-centre flux of the Aq-A1 halo as a function of angular
   separation from the halo centre. The crosses mark the median
   sightline flux for each distribution. The 95 per cent distributions
   are shown as solid lines and continued to 99 per cent as dashed
   lines. The dotted curve is fit to the median data and has the form
   $b^{n}[\phi+b]^{-n}$ where $b = 4.3\,^{\circ}$ and $n = 1.1$.}
   \label{FluxAng}
 \end{figure}

 To begin, we generate 5000 `on-centre' observations of the Aq-A1 halo
 as discussed in subsection~\ref{SSsl}, at a distance of 8kpc from the
 halo centre. In order to generate off-centre observations, we retain the positions used for the on-centre observers and target our
 sightlines in a random direction of some angular separation from the halo
 centre for 5000 sightlines; we do not attempt to identify or exclude
 sightlines that contain large substructures. We perform this procedure for a series of
 angular separations between 10 and 100 degrees in the Aq-A1
 dataset. We take the ratio of each observer's off-centre flux
 measurement and on-centre measurements and plot the results as a
 function of the offset angle, $\phi$, in Figure~\ref{FluxAng}.

The flux ratios taper off with offset angle in a way that can be
approximated by a power
law. We impose a functional form of $b^{n}[\phi+b]^{-n}$ to the data,
thus ensuring that the function has a value of 1 at 0 degrees. Our best
fit for the median data is $b=4.3\,^{\circ}$ and $n=0.94$. We repeated the
fitting procedure for the 95 per cent upper and lower bounds and found that
the best fit parameters when using the same function were $b=4.5\,^{\circ},n=0.71$ and
$b=4.1\,^{\circ},n=1.1$ respectively.

\subsection{Blank sky observations}
\begin{figure}
   \includegraphics[scale=0.35,angle=-90]{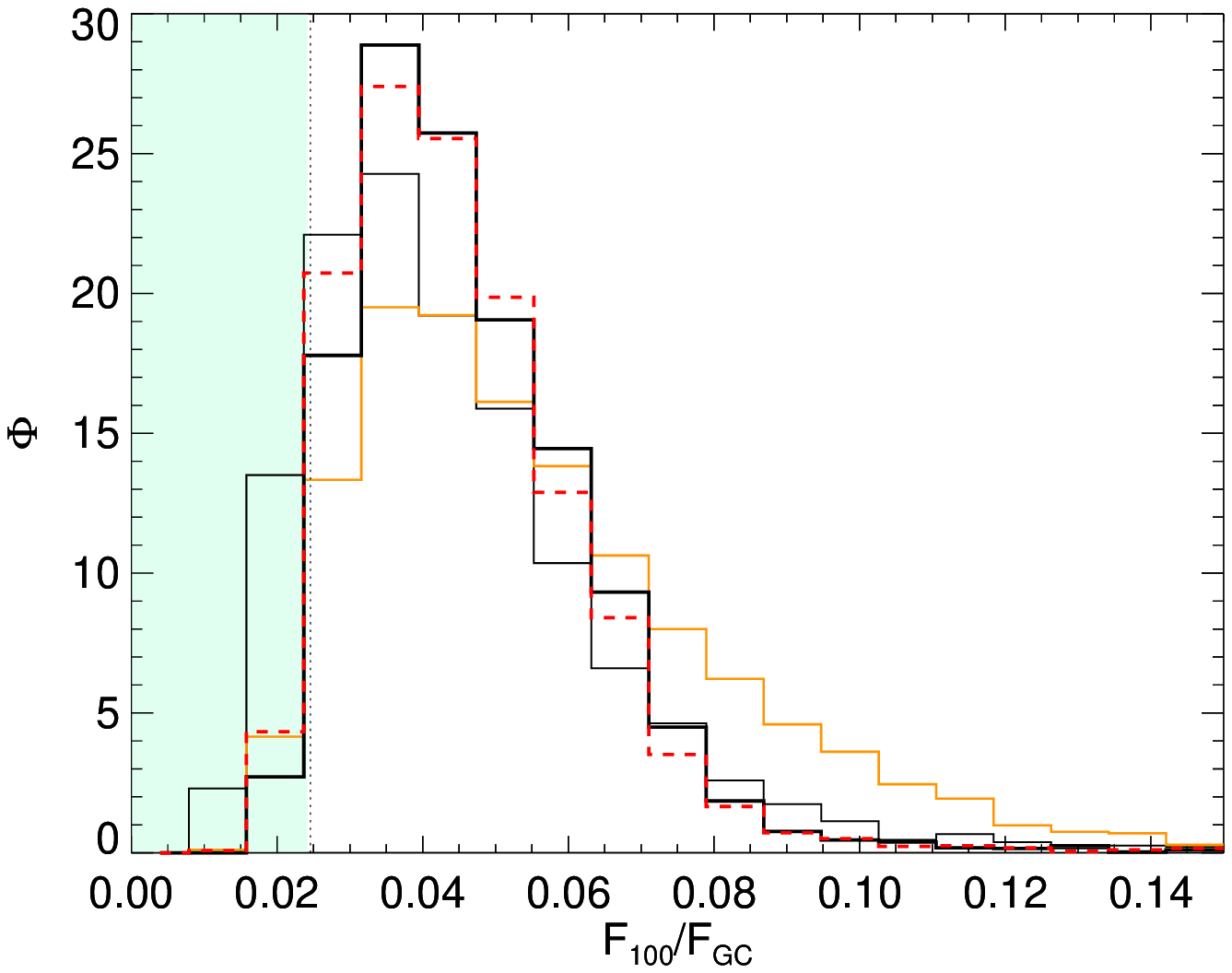}  
   \caption{Histogram of the flux ratios for the $100^{\circ}$ Milky Way
   halo offset angle (black lines, Aq-A1 thick, Aq-A2 thin). We also plot the same result
   for Aq-B2 (orange), and again for Aq-A1 when observers are
   constrained to the plane normal to the minor axis vector (red
   dashed line). The vertical dotted line marks the 95 per cent lower
   bound on the flux distribution. The combined allowed region from
   the $2\sigma$ limit on the \citet{Boyarsky14b} GC detection and
   the \citet{Boyarsky14} blank sky non-detection is given by the
   shaded green region.}
   \label{ROff100Hist}
 \end{figure}

 One particular application of off-centre measurements is the ability
 to obtain a blank sky dataset. \citet{Boyarsky14} used a stack of
 observations taken at
 several offset angles, for simplicity we use $100^{\circ}$. In the same
 way as for Figure~\ref{FluxAng} we calculate the ratio of the $100^{\circ}$
 offset and on-centre measurements. We then bin up these ratios and plot the results in
 Figure~\ref{ROff100Hist} for Aq-A1, Aq-A2, and Aq-B2. We retrieve a
 broad distribution centred
 around 0.05 and obtain a very small probability density either below
 0.02 or above 0.13. The 95 per cent lower bound of the ratio, as
 shown in the Figure, is 0.025. \citet{Boyarsky14} obtained a $2\sigma$ upper
 limit on the blank sky dataset flux of
 $0.7\times10^{-6}\rmn{cts/sec/cm}^{2}$; we would therefore expect the
 flux of the GC to be no higher than
 $\sim3.6\times10^{-5}\rmn{cts/sec/cm}^{2}$ since it can be shown that
 the highest possible value of the GC flux would be the lowest value
 of this ratio multiplied by the $2\sigma$ error on the blank sky
 flux. The Aq-A2 curve differs somewhat from that of Aq-A1, which shows
 that we have not achieved good convergence at least for Aq-A2. We also
 include the result for Aq-B2, as this is the lightest halo and
 therefore perhaps the best GC candidate (albeit with a low
 concentration). When we combine the
 $2\sigma$ exclusion limit from \citet{Boyarsky14} with the reported
 detection in the GC
 \citep[$2.9^{+0.5}_{-0.5}\times10^{-5}\rmn{cts/sec/cm}^{2}$;][]{Boyarsky14b}
 we obtain the shaded region shown in Figure~\ref{ROff100Hist}. We also show the plot
 for observers in the plane of the inner halo minor axis, since this
 is the most likely orientation of the stellar disc with respect to
 the dark matter halo \citep[][see appendix subsection~\ref{PoO}]{Bailin05,Aumer13}. The
 distribution becomes skewed towards lower ratio values, however the
 lower limit to the distribution remains remarkably similar. There is
 some tension between the simulation result and the observational
 constraint. However, it should be noted that we do not attempt to
 identify sightlines that would be the most appropriate matches to
 blank sky targets such as the Lockman hole. Such sightlines may well
 have a lower projected mass density than that the unbiased selection
 offered here: actively selecting underdense light cones could well
 alleviate this tension.

 \subsection{M31 vs. GC observations}
\begin{figure}
   \includegraphics[scale=0.35,angle=-90]{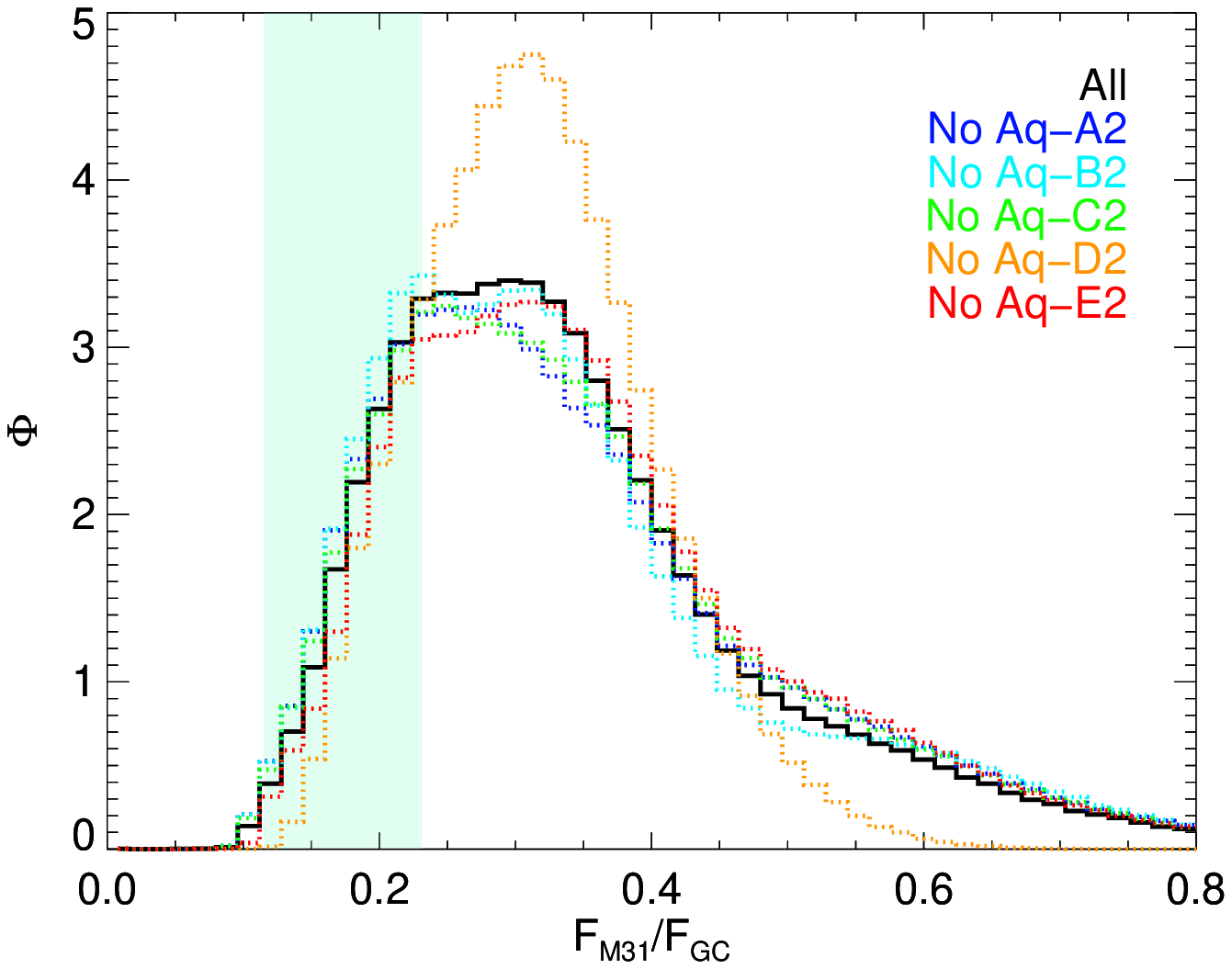}  
   \caption{The probability distribution function for the M31/GC flux
     ratio. The result returned when all haloes are included is shown as
     the solid black line. The dotted lines correspond to cases where the
     data from one halo is omitted from the sample: Aq-A (blue), Aq-B (cyan), Aq-C
     (green), Aq-D (orange) and Aq-E(red). The shaded region
     corresponds to the $1\sigma$ uncertainty on the ratio of the
     detections of M31 \citet{Boyarsky14} and the GC \citet{Boyarsky14b}.}
   \label{RatioNoX}
 \end{figure}

 We now determine the likely ratio of the GC and M31 flux
 measurements. We assume initially that each of the five level 2
 haloes used in our study (Aq-A to Aq-E) are
 all equally likely host haloes for the Milky Way and M31. We combine
 the lists of GC sightlines for each of the five haloes into
 one large catalogue, and likewise for the M31 sightlines. The simulations
 we use are resolution level 2; we multiply the flux of each GC
 sightline by a factor of 1.2 to compensate for resolution
 suppression as derived in Appendix~\ref{RT}. We then extract one
 million randomly selected (with replacement, such that we can pick
 the same sightlines to be part of more than one pair) M31/GC sightline pairs
 in order to build a probability distribution function as the ratio of
 M31/GC observed flux. We plot the results for the combined list of sightlines in
 Figure~\ref{RatioNoX}.

 We find that the distribution peaks at $F_{\rmn{M31}}/F_{\rmn{GC}}=0.3$. The distribution as a whole is
 very broad: 95 per cent of the
 data is in the range [0.15,0.72]. To test how sensitive the result is
 to the properties of individual haloes, in the same Figure we also
 plot the same quantity whilst removing one halo at a time from the sample (as
 both a GC and M31 candidate). The position of the peak changes very
 little as a result of this procedure. However, the omission of the
 Aq-D halo from the sample is seen to remove much of the probability
 distribution at the high and low value tails: the 95 per cent bounds then
 tighten to
 [0.18,0.52]. The effect of removing any of the other four haloes
 is much smaller by comparison. The position of the peak varies
 between 0.23 and 0.33.   

 \citet{Boyarsky14,Boyarsky14b} claim a detection from M31 of a
 line with flux
 $4.9^{+1.6}_{-1.3}\times10^{-6}\rmn{cts/sec/cm}^{2}$, and in the GC
 of  $2.9^{+0.5}_{-0.5}\times10^{-5}\rmn{cts/sec/cm}^{2}$. Combining
 these two results and their associated errors, we obtain the shaded
 region shown in Figure~\ref{RatioNoX}. It is in broad agreement with
 all combinations of our haloes.

 \begin{figure}
   \includegraphics[scale=0.35,angle=-90]{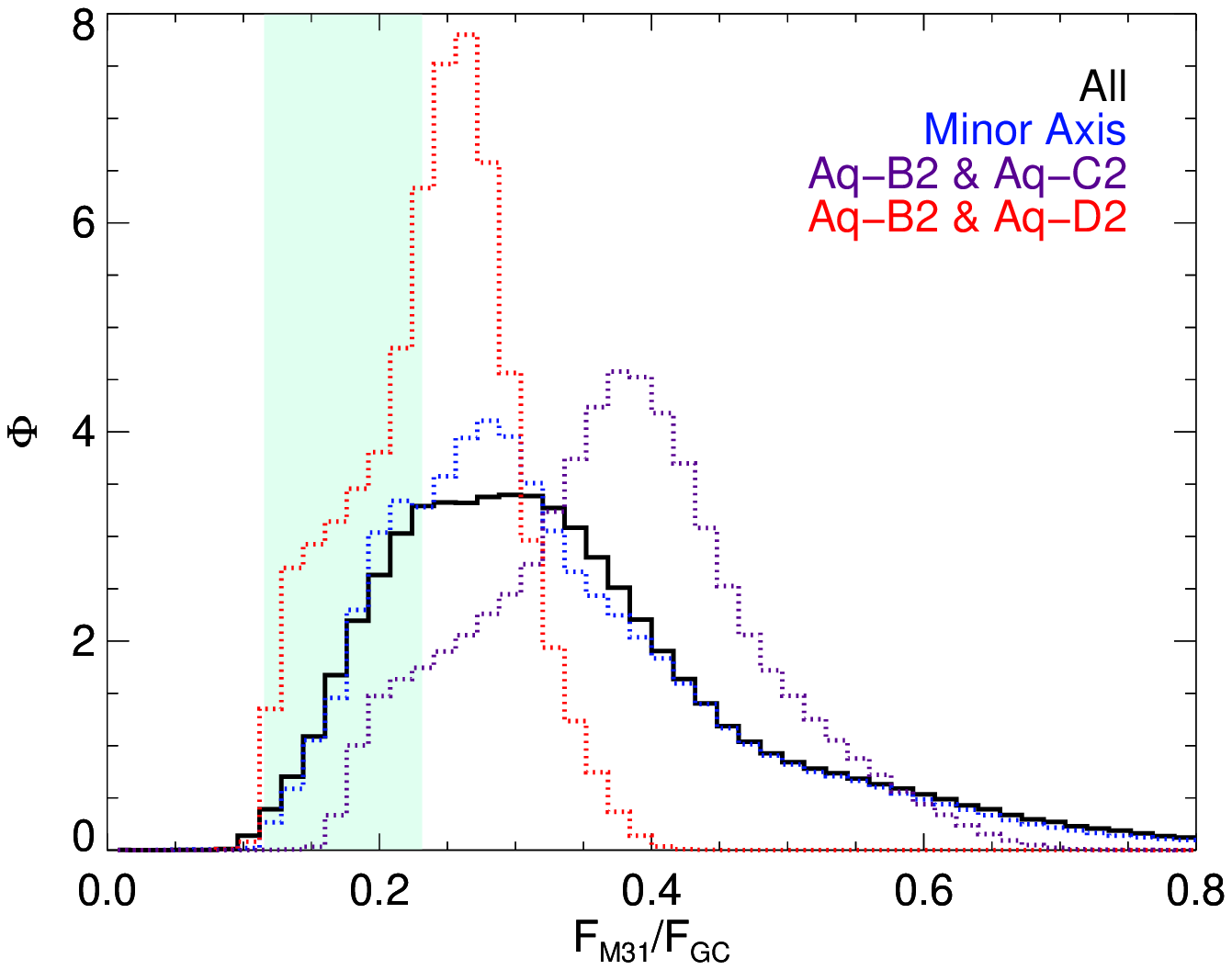}  
   \caption{The probability distribution function for the M31/GC flux
     ratio is reproduced in black (all haloes). We also include data
     for observers in the plane of the minor axis (blue dotted
     line). The other two lines are obtained when our GC target is Aq-B2 and
     the M31 halo analogue is taken to be Aq-C2 (purple) and Aq-D2
     (red). We reproduce the observational constraints as the shaded
     green region.}
   \label{RatioMassMinor}
 \end{figure}

 \medskip

 We now consider other variables that may affect the distribution
 of M31/GC flux ratios. In Figure~\ref{RatioMassMinor} we plot three
 additional versions of the flux ratio histogram. The first is for GC
 observers constrained to reside in the inner-halo minor axis plane. The requirement that the
 observers reside in the minor axis plane has very little effect on
 the result. The second and third consider the
 case in which M31 is much more massive than the Milky Way
 \citep{Penarrubia14}. We take Aq-B2 -- our smallest halo -- to be our
 Milky Way candidate and Aq-C2 and Aq-D2 to be M31 halo candidates. Despite
 having the same mass to four significant figures, the flux histograms for the
 two combinations of GC and M31 (Aq-B2 -- Aq-C2 vs. Aq-B2 -- Aq-D2)
 are displaced by a factor of 1.6; the large difference in
 concentration (18.4 for Aq-C2, 12.4 for Aq-D2) plays an important role in the result.  

 \subsection{Dwarf spheroidal galaxies}
 Additionally, one can hope to detect the signal in still more targets. Dwarf
 spheroidal galaxies are exceptionally good objects for further
 study~\citep{Boyarsky:06c,Riemer:09a,Malyshev14}: they have very high mass-to-light ratios
 \citep[e.g.][]{Wolf10}, they represent a different mass regime to clusters
 and $L_{*}$ galaxies, and also contain very little if any interstellar gas
 \citep[see][and references therein]{Gallagher03}. They therefore offer a set
 of circumstances in which the ratio of the predicted dark matter emission to
 the astrophysical background is very high.

  The mass enclosed in the half-mass radius has
  been estimated by several studies \citep{Walker09, Walker10,
  Wolf10}. Two of the satellites measured to
 have the highest central densities are Draco and Sculptor
  \citep{GeringerSameth2014}. We will
 select Draco and Sculptor candidates from our simulations and use
 these to make predictions for the likely amplitude of X-ray decay
 fluxes from these two satellites. We reproduce the observational data
 published in \citet{Wolf10} in Table~\ref{TabDS}.

 \begin{table}
    \centering
    \begin{tabular}{|l|l|l|}
      \hline
       & Draco & Sculptor \\
      \hline
       $d$ (kpc) & $76\pm5$ & $86\pm5$ \\
       $L$ ($L_{\odot,V}$) & $2.2^{+0.7}_{-0.6}\times10^{5}$ & $2.5^{+0.9}_{-0.7}\times10^{6}$\\
       $r_{1/2}$ (pc) & $291\pm14$ & $375\pm54$ \\
       $M_{1/2}$ ($\Msun$) & $2.11^{+0.31}_{-0.31}\times10^{7}$ &
       $2.25^{+0.16}_{-0.15}\times10^{7}$ \\       

      \hline
    \end{tabular}
    \caption{Selected parameters of the the Draco and Sculptor dwarf
    spheroidals as reproduced from \citet{Wolf10}: distance, $d$,
    luminosity, $L$, de-projected 3D half light radius, $r_{1/2}$, and
    half light mass, $M_{1/2}$.}
    \label{TabDS}
  \end{table}

 We select Draco and Sculptor candidate subhaloes as follows. For each
 of our level 2 simulations we identify subhaloes that are between 88
 and 148~kpc from the main halo centre. These values are chosen to
 increase our sample size beyond what is possible at the true distance
 of these satellites, such that the observer will be at a distance of
 80-140kpc from the satellite. We then draw a sphere of radius
 equal to the Draco / Sculptor half-light radii around the subhalo
 centre. If the mass
 enclosed within that radius falls within the published uncertainty on
 the `half-light mass' of the satellite galaxy in question then it is
 added to our sample. In this way we obtain 19 candidates for Draco
 and 33 for Sculptor. The shapes of the density profiles of these
 simulated subhaloes is quite different to that inferred for dwarf
 spheroidals by some studies (c.f. \citealp{Springel08b,
 Walker11} but see also \citealp{Strigari14}) however our
 study is sensitive to the total subhalo mass within the satellite half-light radius alone, and not the density profile.

\begin{figure}
   \includegraphics[scale=0.35,angle=-90]{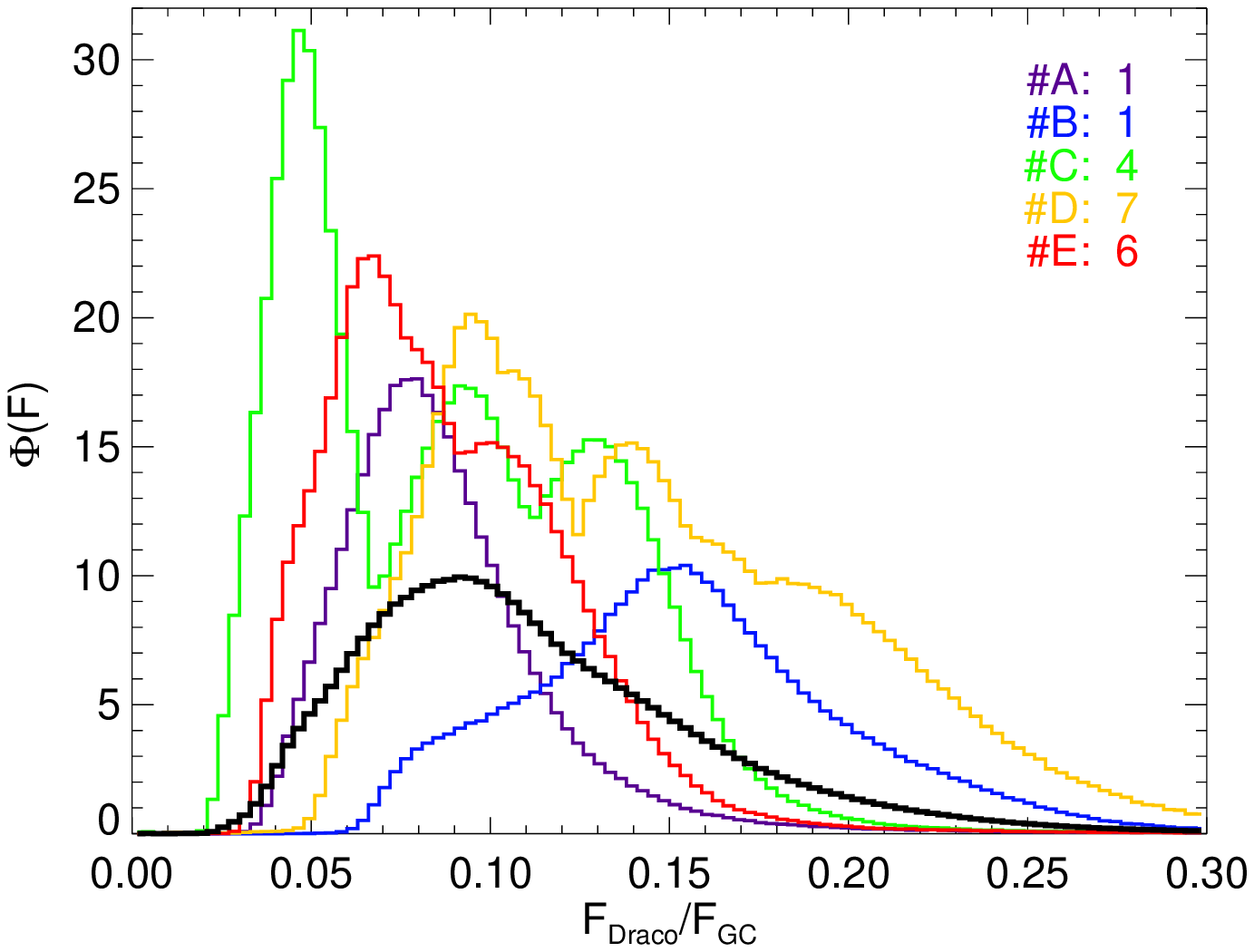} 
   \includegraphics[scale=0.35,angle=-90]{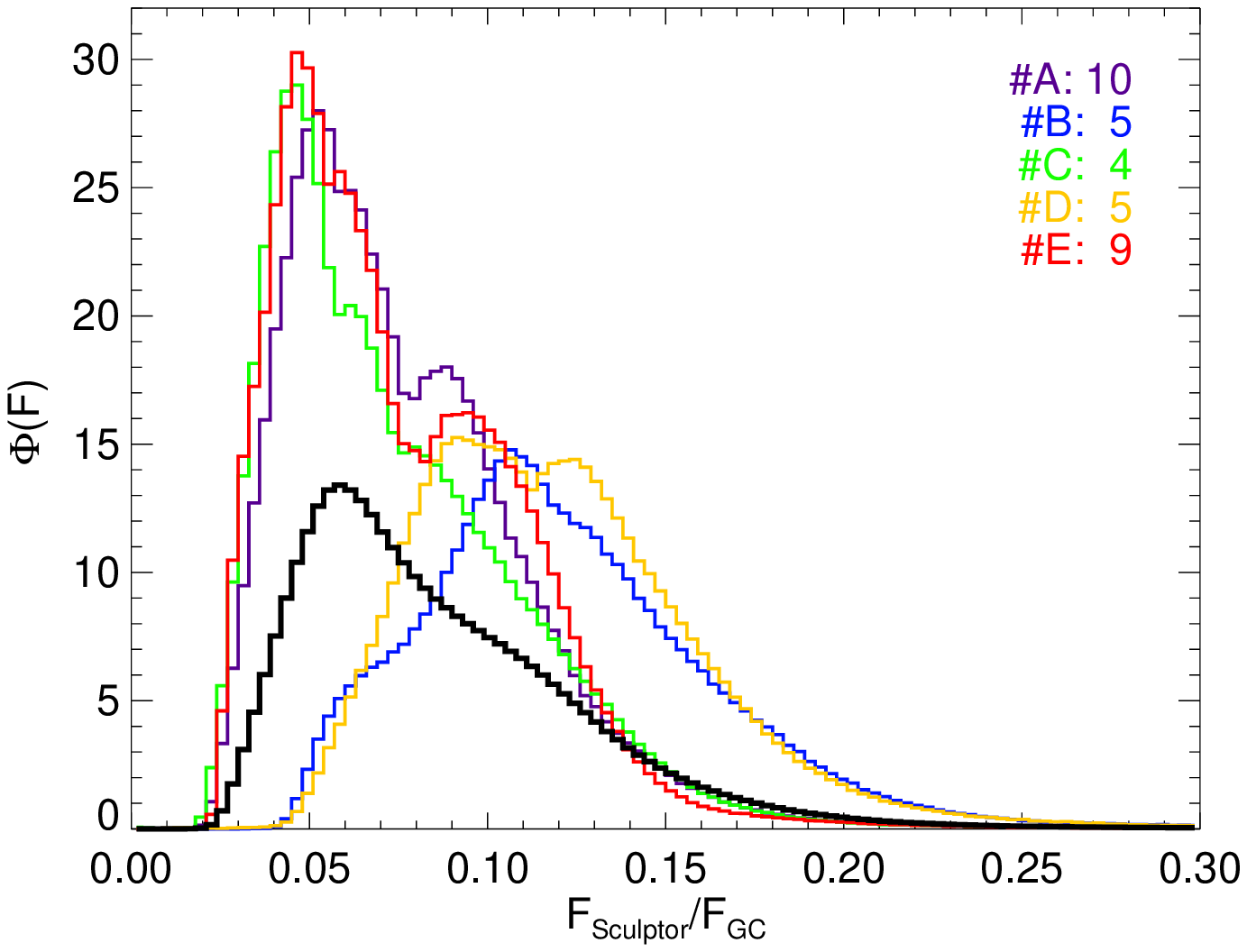}  
   \caption{The probability distribution function envelopes for the flux ratios
   of Draco/GC (top panel) and Sculptor/GC (bottom panel). We
   calculate the normalised histogram for each individual satellite/GC
   pair and then, for each central halo, plot the curve that envelopes
   all of the curves associated with that central. The new histogram
   is not renormalised. The histograms for central haloes Aq-A2, Aq-B2, Aq-C2, Aq-D2, and Aq-E2 are shown in
   purple, blue, green, orange, and red respectively. The black line
   is the distribution obtained when all the distributions in each
   panel are merged.}
   \label{DraScu}
 \end{figure}

 We place 5000 observers at random within the ring defined such that the
 observer-main halo distance is 8kpc and the observer-target subhalo
 distance is either 80, 90, 100, 110, 120, 130, or 140kpc from the
 centre as permitted by the main halo-subhalo separation. For subhalo
 positions such that more than one of these observer-target
 separations are possible we select the smallest available. We then
 calculate the flux from the target at the position of the observers
 in the same way as for the GC and M31. By this method we obtain the
 combined flux from the subhalo and the outskirts of the main
 halo. The results for both Draco and Sculptor are presented in Figure~\ref{DraScu}.

 \medskip

There is a great deal of variation between different dwarf spheroidal
candidates. For Draco the peak in the probability distribution may be as much as
20 per cent of the GC or as little as 5 per cent; for
Sculptor the range in peaks is smaller. It it possible to see the
effects of central halo concentration and mass: the Aq-D2 results are all
clustered towards more modest ratios, as are those of Aq-B2.  

\subsection{Compatibility of X-ray line claims and limits in different targets}

\begin{figure*}
   \includegraphics[scale=0.55,angle=-90]{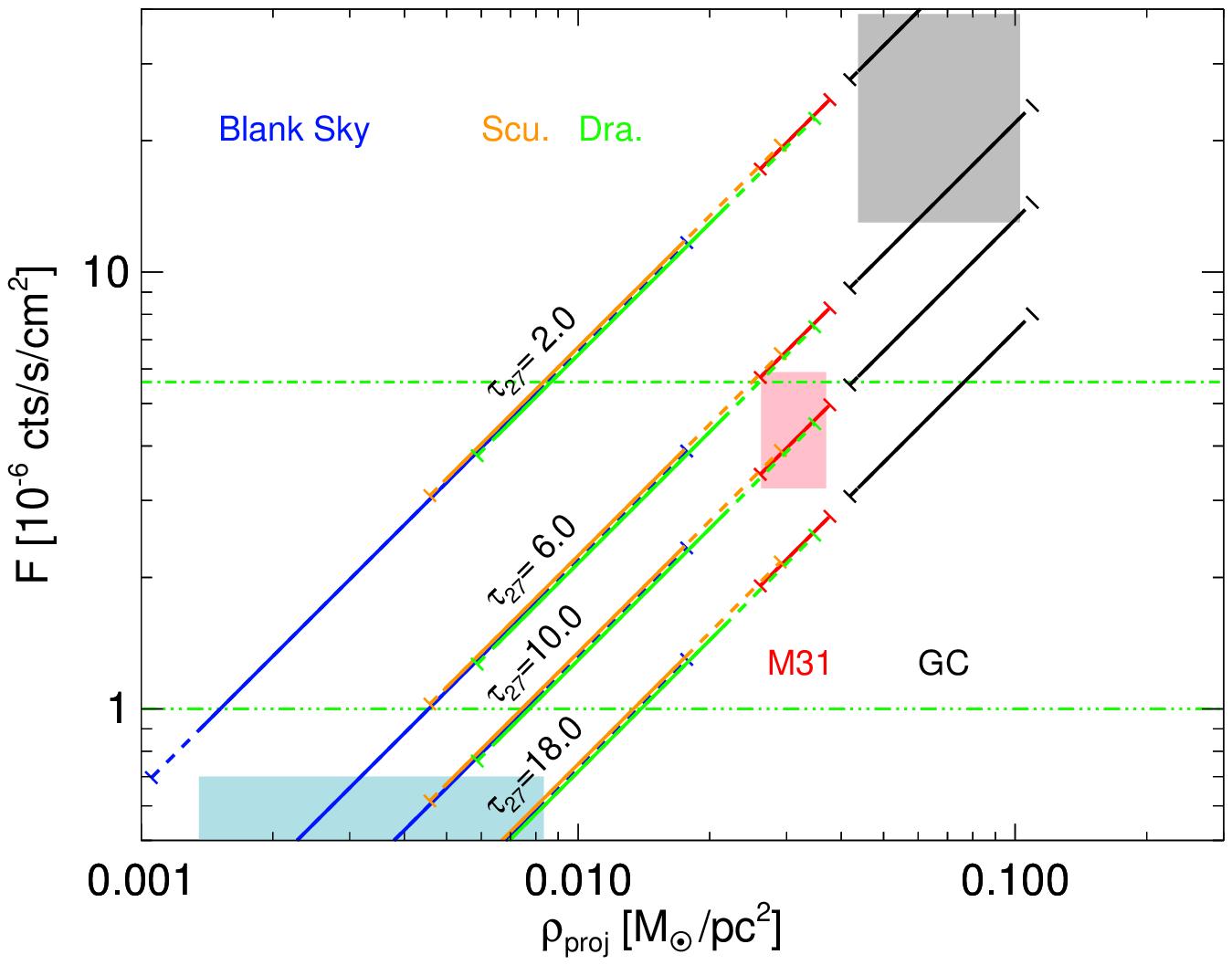}    
   \caption{The distribution of fluxes from each of our targets as a
   function of the projected mass density for a series of four decay
   lifetimes ($\tau_{27}=2,6,10,18$). The GC (Aq-B2), M31 (Aq-D2),
   Draco (all candidates), Sculptor (all candidates) and Milky Way
   $100^{\circ}$ offset (Aq-B2) are denoted by the black, red, green,
   orange, and blue lines respectively. Lines are solid for the 95 per
   cent region and extended to 99 per cent with dashed lines. The
   grey shaded region denotes the $1\sigma$ allowed region from the GC
   detection of \citet{Boyarsky14b}, the red shaded region that of
   the M31 detection \citep{Boyarsky14}, and the blue shaded region
   the $2\sigma$ upper bound from the \citet{Boyarsky14} blank sky
   dataset. The dot-dashed green line represents the current $2\sigma$
   upper bound on the flux from Draco (107.1~ks of XMM-MOS1 and XMM-MOS2
   data; 40.4~ks XMM-PN data) and the triple-dot-dashed green line
   the $3\sigma$ limit that would be expected for $\sim1.3$~Ms of XMM
   data. Good agreement between
   the simulation predictions and the observations is achieved when
   the lines overlap with their associated shaded regions for a
   single value of $\tau_{27}$.}
   \label{LFlux}
 \end{figure*}

We now bring the results from each of these targets together to
examine the likelihood that each (non-)detection is consistent with
being generated by a dark matter particle of the same decay
lifetime. In Figure~\ref{LFlux} we plot the flux distributions for
each of our targets as a function of projected mass for a series of
different decay lifetimes $\tau_{27}=\tau/(10^{27}\rmn{s})$. We take Aq-B2 to
be our Milky Way candidate due to its low mass \citep{Deason12,Penarrubia14}, and Aq-D2 to be our
M31 analogue because of its relatively low concentration\citep{Corbelli10}. We also include
the $1\sigma$ allowed regions for the detections of the GC and M31, and
also $2\sigma$ upper bounds for the blank sky dataset and Draco.

 We find that the best agreement between the corresponding simulation
 predictions and the observational constraints is approximately in the range
 $\tau_{27}=(6-10)$. This lifetime agrees within $1\sigma$ range with the
 results of~\citet{Bulbul14,Boyarsky14,Boyarsky14b} and is consistent with
 non-observation of the line from stacked observations of dSphs
 \citep{Malyshev14} where the limit $\tau_{27} > 7.3$ ($3\sigma$ upper bound)
 was established. There is more tension with the limits from
 \citet{Anderson14}, who claim to rule out the line as found by
 \citep{Bulbul14} at as much as $11.8\sigma$, however their
 applications of scaling relations and their stacking procedure may have consequences for the error estimates.
 These estimates are based on our haloes that are the
 best candidate matches for the Milky Way and GC. The projected mass
 will increase if the expected mass or concentration of either target
 were higher will cause the projected mass to change slightly. A
 proper implementation of baryon physics in particular will likely
 increases the projected mass densities (see Appendix~\ref{MI}).

 In addition to the published upper bound from the blank sky observations and
 also the published detections in the GC and M31, we include in
 Figure~\ref{LFlux} the upper limit from the non-detection of Draco, as well
 as an estimate of the sensitivity of XMM-Newton with future
 observations. Based on archival data from the XMM-Newton observatory
 ($\sim107$ks of MOS1+MOS2 instruments, and 40.4ks with PN) the $2\sigma$
 upper bound is $5.6\times10^{-6}\rmn{cts/s/cm}^{2}$. Simulations using the
 Xspec's~\citep{Arnaud:96} \textsc{fakeit} command find that a Draco flux of
 $\sim1\times10^{-6}\rmn{cts/sec/cm}^{2}$ would be detected at $3\sigma$ with
 an XMM-Newton exposure of 1.34~Ms. For values of $\tau_{27} \sim 8$, 95 per
 cent of our realizations lead to an X-ray flux from Draco consistent with
 existing upper limits, and within the 3$\sigma$ reach of XMM-Newton with an
 exposure of 1.34~Ms.
 
 \section{Conclusions}
 \label{Conc}

 The identity of the dark matter remains unknown. The detection of a
 series of unexplained lines in the X-ray spectra of a number of
 different astrophysical targets are suggested to be consistent with
 the decay of light dark matter particles. We have expanded on these
 studies by estimating the likely distribution
 of X-ray fluxes from these targets. 

 To this end we have
 used simulations of Milky Way-analogue dark matter haloes to ascertain
 the likely signal of the decay of dark matter into
 X-ray photons from the Galactic centre, M31 and two dwarf spheroidal
 galaxies. We placed 5000 observers at a distance of 8kpc
 around the simulated main halo centre and calculated the flux
 measured within the FoV of radius $14'$, treating each
 simulation dark matter particle as a point source of X-ray
 photons. This procedure were also performed with observers
 780kpc from the halo centre to simulate the likely M31 flux.  

  
 We then calculated the likely signal from other parts of the Milky
 Way halo. We performed an off-centre analysis in
 which we take a series of 5000 sightlines $100\,^{\circ}$ away from
 the main halo centre. We obtain a flux distribution that is slightly
 in tension with the 95 per cent exclusion limit from the \citet{Boyarsky14}
 blank sky dataset.

 Pairs of GC and M31 measurements were compared to the ratio of
 fluxes found by \citet{Boyarsky14,Boyarsky14b}, and were found to be in
 good agreement for a wide range of sampling procedures. 

 Finally, we repeated the procedure for the Draco and Sculptor dwarf
 spheroidal satellite galaxies. We find a wide range of possible
 probability curves, with the central halo density as a very
 important parameter. Our results in this regime are consistent with a
 non-detection reported using $\sim100$ks of archival {\it XMM-Newton}
 data, and future pointings of Draco will have the ability to rule out a dark matter decay
 signal from this object for values of $\tau_{27}<8$ at 95 per cent confidence. 

 \section*{Acknowledgements}

 We would like to thank Dima Iakubovskyi and Jeroen Franse for reading the
 manuscript and for collaboration on X-ray data analysis and on estimates of
 sensitivity of future observations. We would also like to thank Christoph
 Weniger for useful comments, and the anonymous referee for a careful reading of the text and helpful suggestions. This work used the DiRAC Data Centric system at
 Durham University, operated by the Institute for Computational Cosmology on
 behalf of the STFC DiRAC HPC Facility (www.dirac.ac.uk). This equipment was
 funded by BIS National E-infrastructure capital grant ST/K00042X/1, STFC
 capital grant ST/H008519/1, and STFC DiRAC Operations grant ST/K003267/1 and
 Durham University. DiRAC is part of the National E-Infrastructure. This work
 is part of the D-ITP consortium, a program of the Netherlands Organisation
 for Scientific Research (NWO) that is funded by the Dutch Ministry of
 Education, Culture and Science (OCW). GB acknowledges support from the
 European Research Council through the ERC Starting Grant WIMPs Kairos.

 \bibliographystyle{mn2e}

 \appendix
  \section{Resolution and Sightline Tests}
  \label{RT}

\begin{figure}
    \includegraphics[scale=0.35,angle=-90]{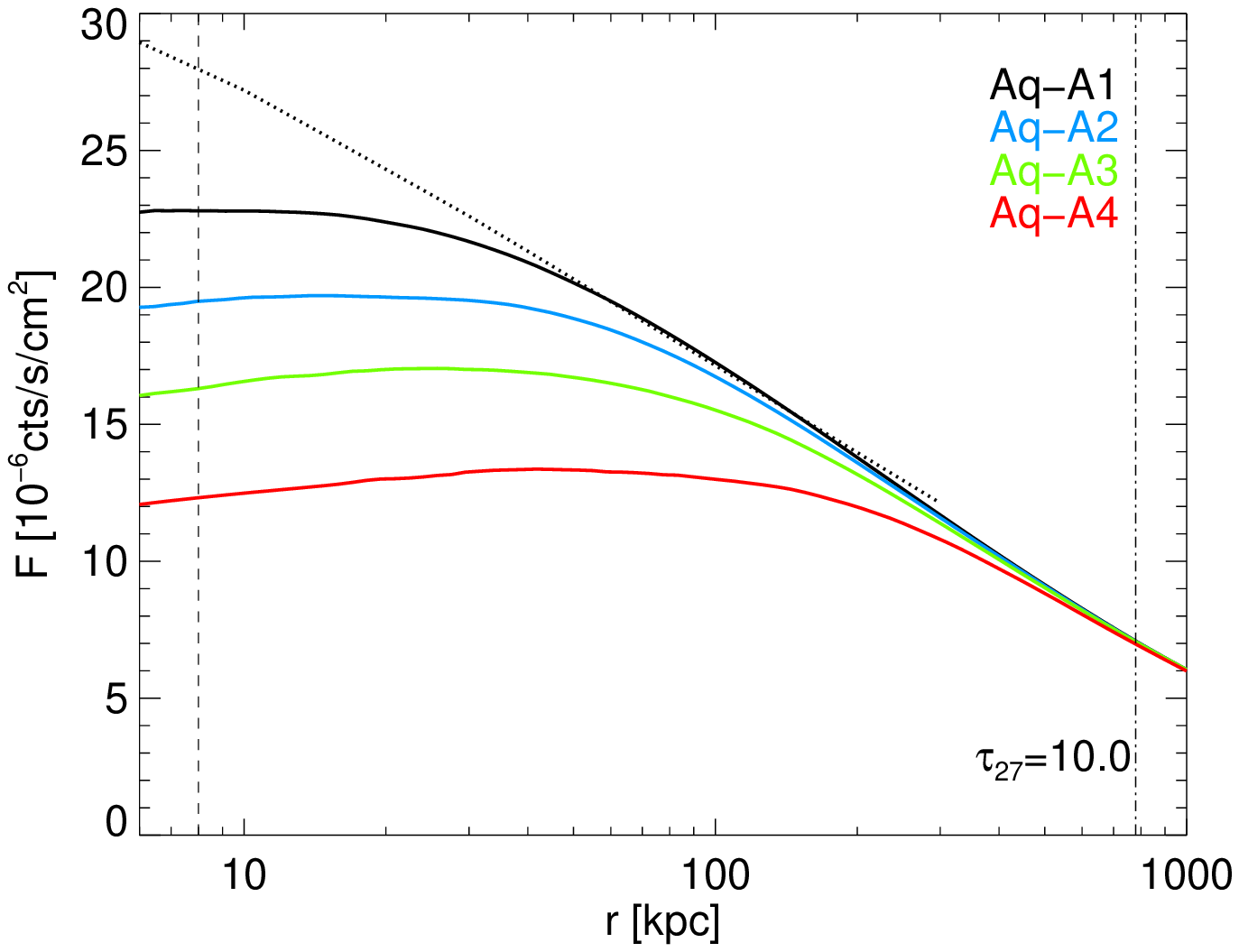}
    \caption{The expected X-ray flux from a Milky Way-analogue DM halo (Aq-A)
      as a function of distance from the halo centre. Each curve
      represents a different resolution simulation: Aq-A1 (black), Aq-A2
      (blue), Aq-A3 (green), and Aq-A4 (red). The dotted line at 780~kpc
      marks the distance to M31, and the dashed line at 8~kpc the
      distance to the GC. We also include an NFW profile with the
      same $M_{200}$ and $c$ as the Aq-A1 halo (dotted line); it is
      normalised to the Aq-A1 curve at 50~kpc.}
    \label{FigSurfaceDens}
  \end{figure}

  We discuss here the convergence with numerical resolution of our simulations. In
  Figure~\ref{FigSurfaceDens} we plot surface density flux as a
  function of radius for a generic decaying dark matter particle of
  decay lifetime $\tau=10^{28}\rmn{s}$, and are not interested in the
  mass of the dark matter particle.  At the distance of M31, the separation between
  all four simulations is minimal and therefore we can consider even
  our level 4 simulation to resolve the system accurately. At smaller
  radii the curves start to diverge systematically with
  resolution. The flux amplitude at the GC distance is suppressed by
  70 per cent in Aq-A4 relative to Aq-A1, compared to a difference in
  particle mass between the two runs of over two orders of
  magnitude. We also plot the expected flux for an Navarro-Frenk-White
  profile \citep[NFW;][]{NFW_96,NFW_97} with the same
  $M_{200}$ and $c$ as our Aq-A1 halo
  ($M_{200}=1.839\times10^{12}\Msun$ and $c=18.6$) for distances
  between 7~kpc and 500~kpc. It predicts approximately 22 per cent more
  flux than our Aq-A1 halo. However, \citet{Navarro10} found that the Aquarius
  haloes were better described by an Einasto profile \citep{Einasto65}
  than by an NFW. The Einasto profile has a shallower central
  slope compared to the NFW, so the NFW curve may be interpreted as
  an upper limit on the `true' flux. Therefore it is unlikely that
  a simulation of infinite resolution would return an GC flux more than $\sim20$ per cent higher than that
  found in Aq-A1.

\begin{figure*}
    \includegraphics[scale=0.35,angle=-90]{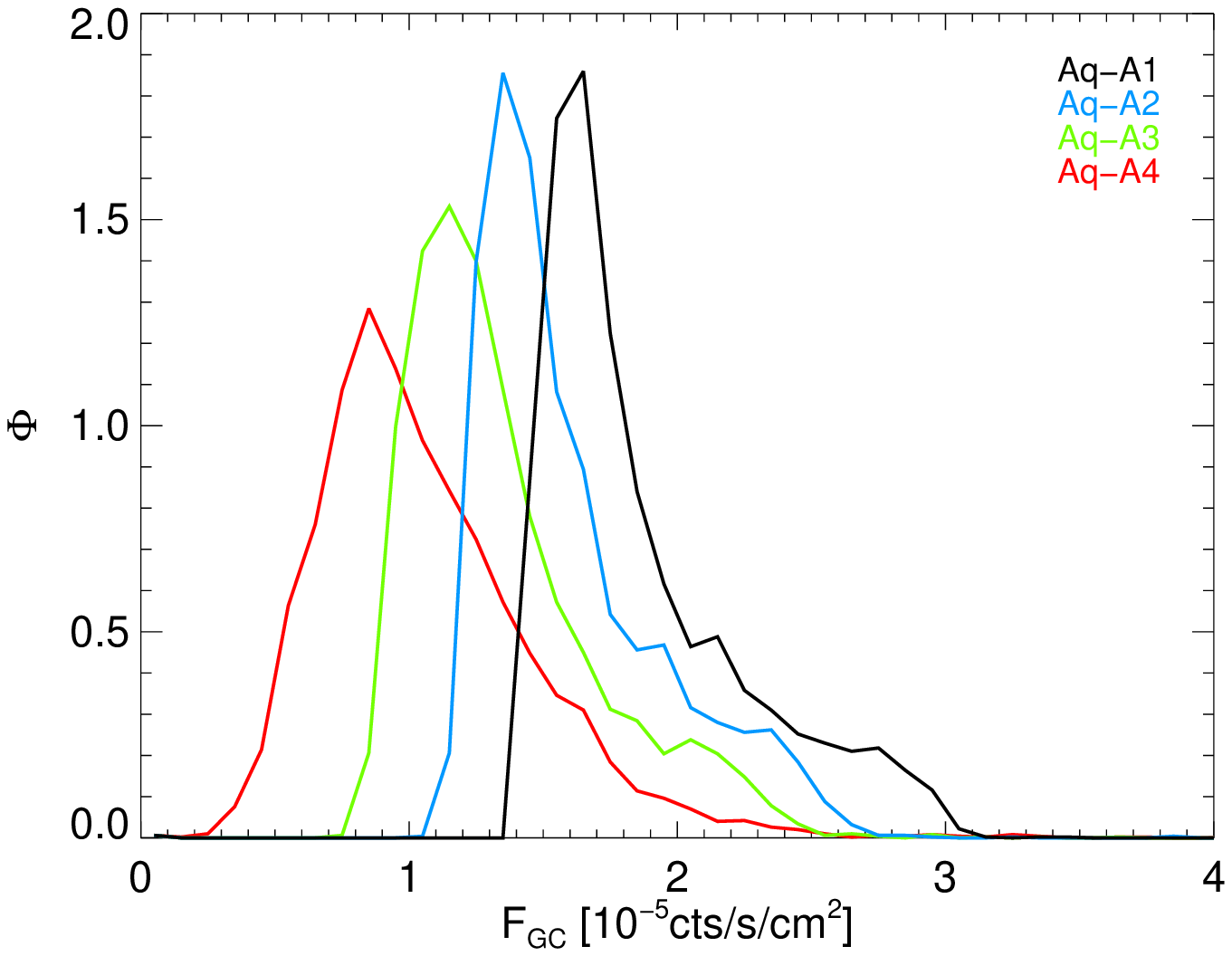}
    \includegraphics[scale=0.35,angle=-90]{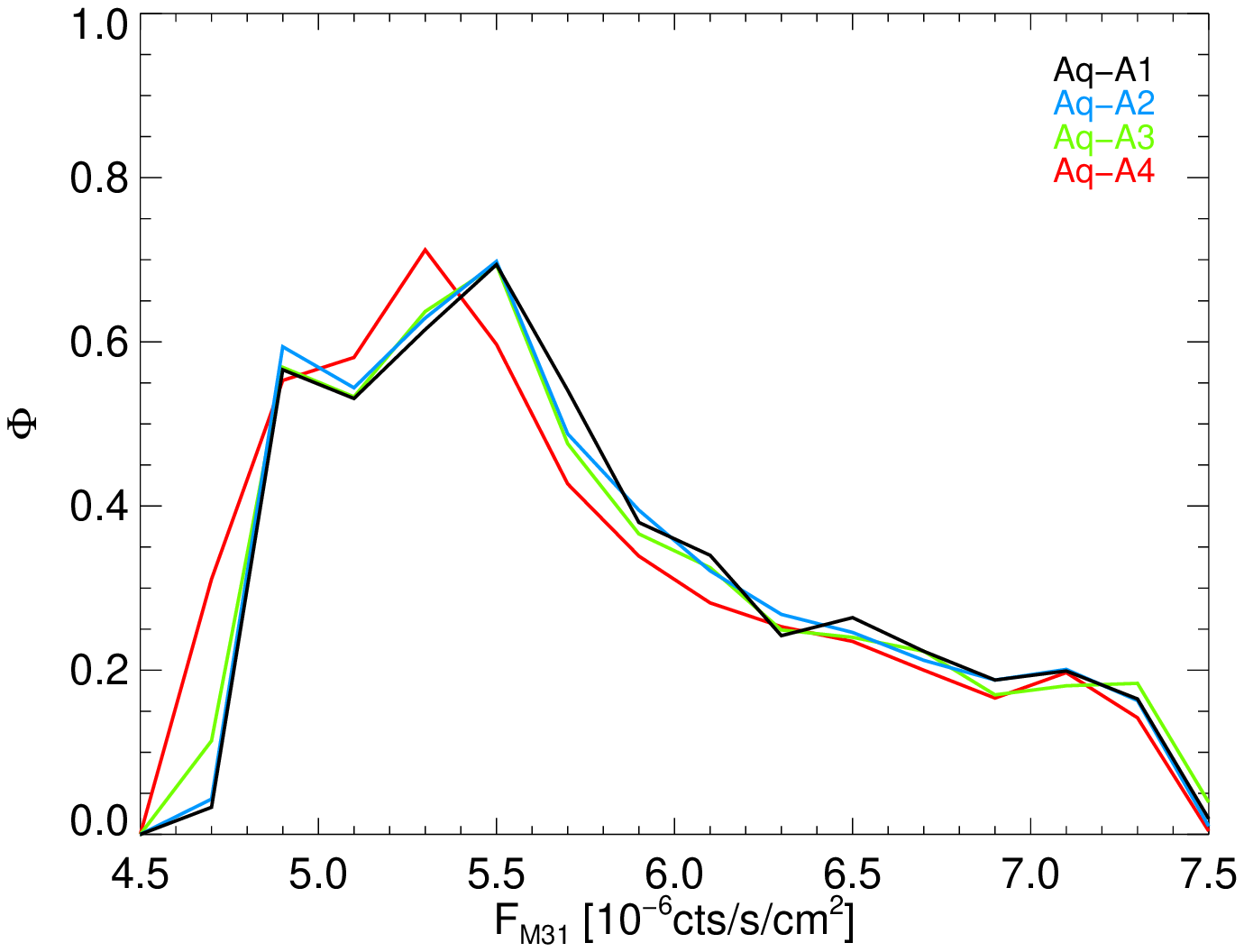}
    \caption{The flux distributions, $\Phi$,  for the GC (left) and M31
      (right) when different resolution simulations of the Aq-A
      halo are used as the target halo. Each distribution is
      normalised such that the area under the curve is equal to 1. The
      fluxes for Aq-A1,
      Aq-A2, Aq-A3, and Aq-A4
      are shown in black, blue, green, and red respectively.}
    \label{FigSLConv}   
  \end{figure*}

  A more demanding criterion for convergence is that our sightline
  measurements converge. In Figure~\ref{FigSLConv} we plot the
  distribution of fluxes for the GC and M31 for different resolution
  simulations of the Aq-A halo. The peak in the GC distribution moves to
  higher fluxes with increasing resolution, since the addition of more
  particles to the simulation increases the mass contained within the
  FoV on small scales. The distribution of the second highest
  resolution -- Aq-A2 -- is suppressed relative to that of the highest
  by 20 per cent, which is comparable to the suppression
  between these two simulations in
  Figure~\ref{FigSurfaceDens}. Interestingly, the shape and width of the
  distribution changes very little between runs Aq-A2 and Aq-A1, which
  suggests that our derived bounds relative to the distribution peak
  for the fluxes will not be affected
  by resolution as much as the amplitude.

  As was the case for the spherically-averaged method, the convergence
  of the M31 measurement is much better. The Aq-A1, Aq-A2, and Aq-A3
  distributions all peak at the same flux and have very similar
  shapes. This is due to the physical radius of the FoV at the target in
  configuration space (740pc) being much larger than the spatial resolution of
  the simulation, unlike the GC case (only 8pc). 

  Finally, we checked that 5000 sightlines measurements are sufficient for
  the convergence of the flux distribution functions and adopt this number
  throughout.

 \section{Influence of cosmology, baryon physics, and halo properties}
 \label{MI}

 In this section we examine the impact of various factors on our
 estimations of the fluxes from the GC and M31,
 including the possible effects of our choice of cosmological
 parameters, warm dark matter power spectra,
 baryons, and halo mass and concentration. 

 \subsection{Cosmological parameters}
\begin{figure*}
   \includegraphics[scale=0.35,angle=-90]{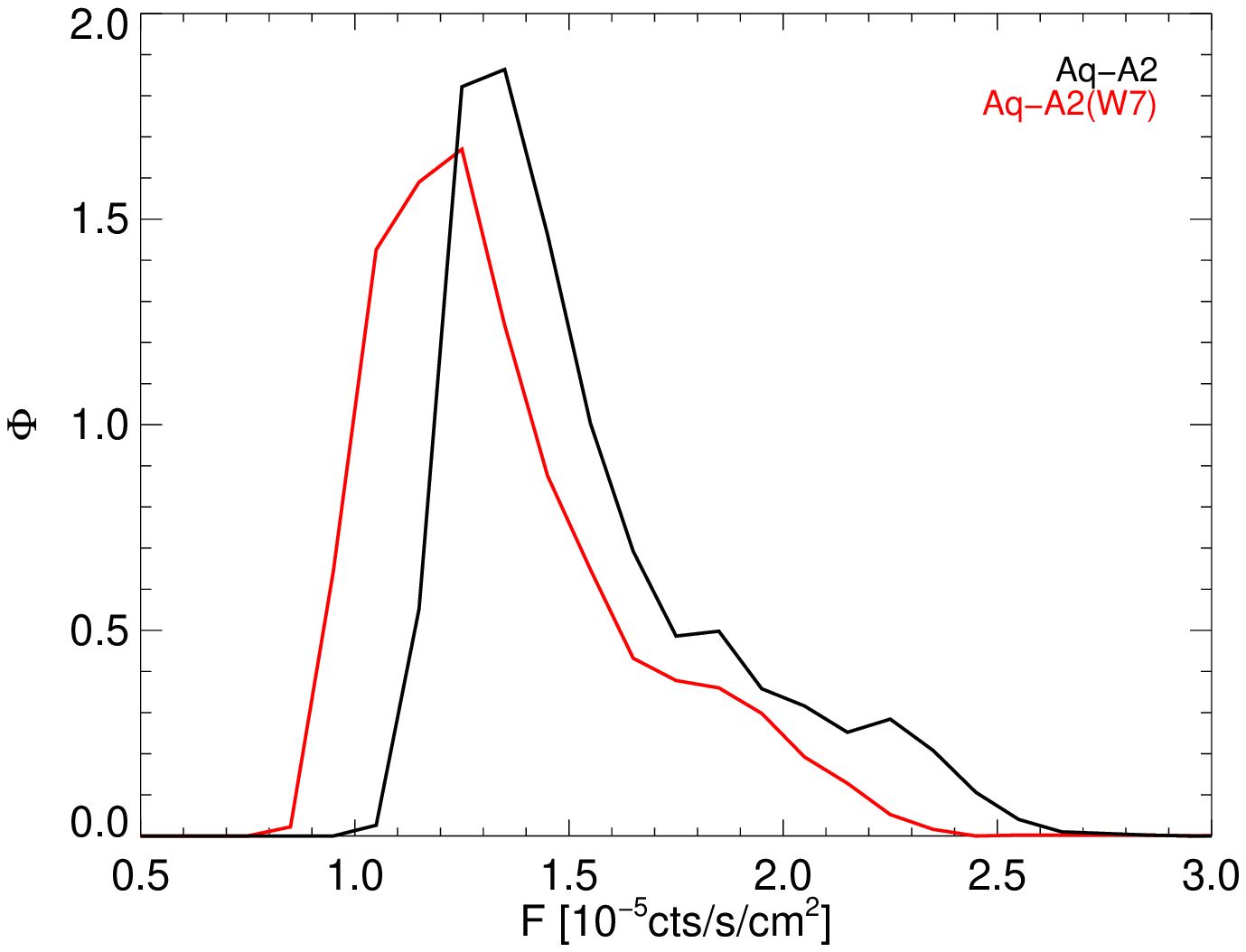}
   \includegraphics[scale=0.35,angle=-90]{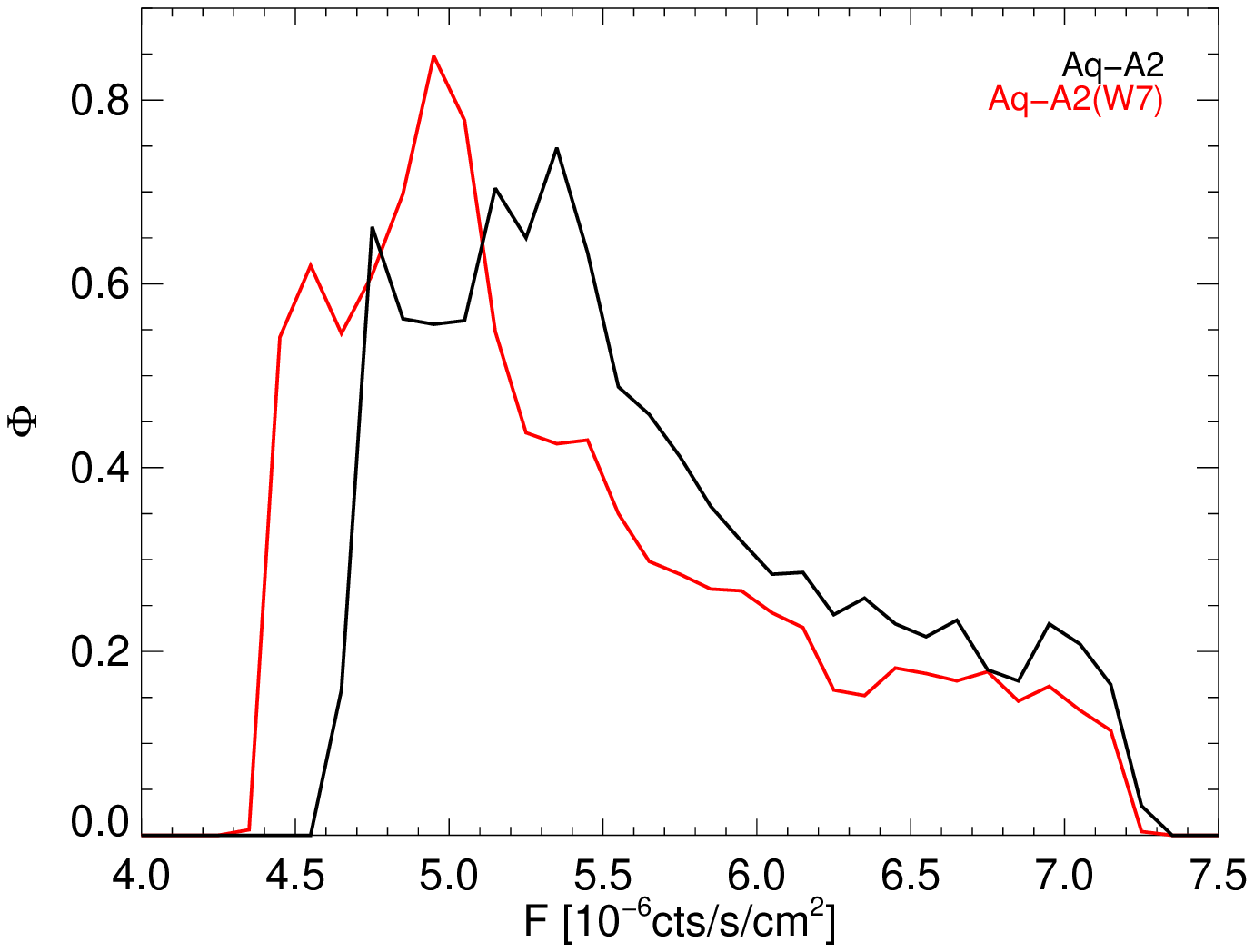}
   \caption{The flux distributions for Aq-A2 (black) and
     Aq-A2(W7) (red) for the GC (left) and M31 (right).}
   \label{FigSLWMAP}
 \end{figure*}

 First we consider the effect of cosmology. The halo properties will
 be sensitive to parameters such as the age of the Universe and at
 what redshift the halo centres form. In Figure~\ref{FigSLWMAP}
 we plot the cumulative flux functions of the {\it WMAP1} and {\it WMAP7} versions of
 Aq-A2. In the GC and M31 regimes the {\it WMAP7} flux distribution is
 suppressed relative to {\it WMAP1} by up to 10 per cent. This occurs despite the
 increase in $M_{200}$ from {\it WMAP1} to {\it WMAP7}; instead the
 halo becomes
 less concentrated. $\sigma_{8}$ has a lower value in {\it WMAP7}
 compared to {\it WMAP1} (see
 table~\ref{TabSim}), and this change delays the halo formation
 time to an epoch when the Universe is less dense \citep{Lovell14,
 Polisensky14}. Any such discrepancy may be magnified artificially by
 resolution issues since the {\it WMAP7} run particle mass is
 slightly higher. In conclusion, it is likely
 that the choice of cosmological parameters will have a impact on the
 M31 and GC fluxes of not more than a few per cent.

 \subsection{Warm dark matter}
\begin{figure*}
   \includegraphics[scale=0.35,angle=-90]{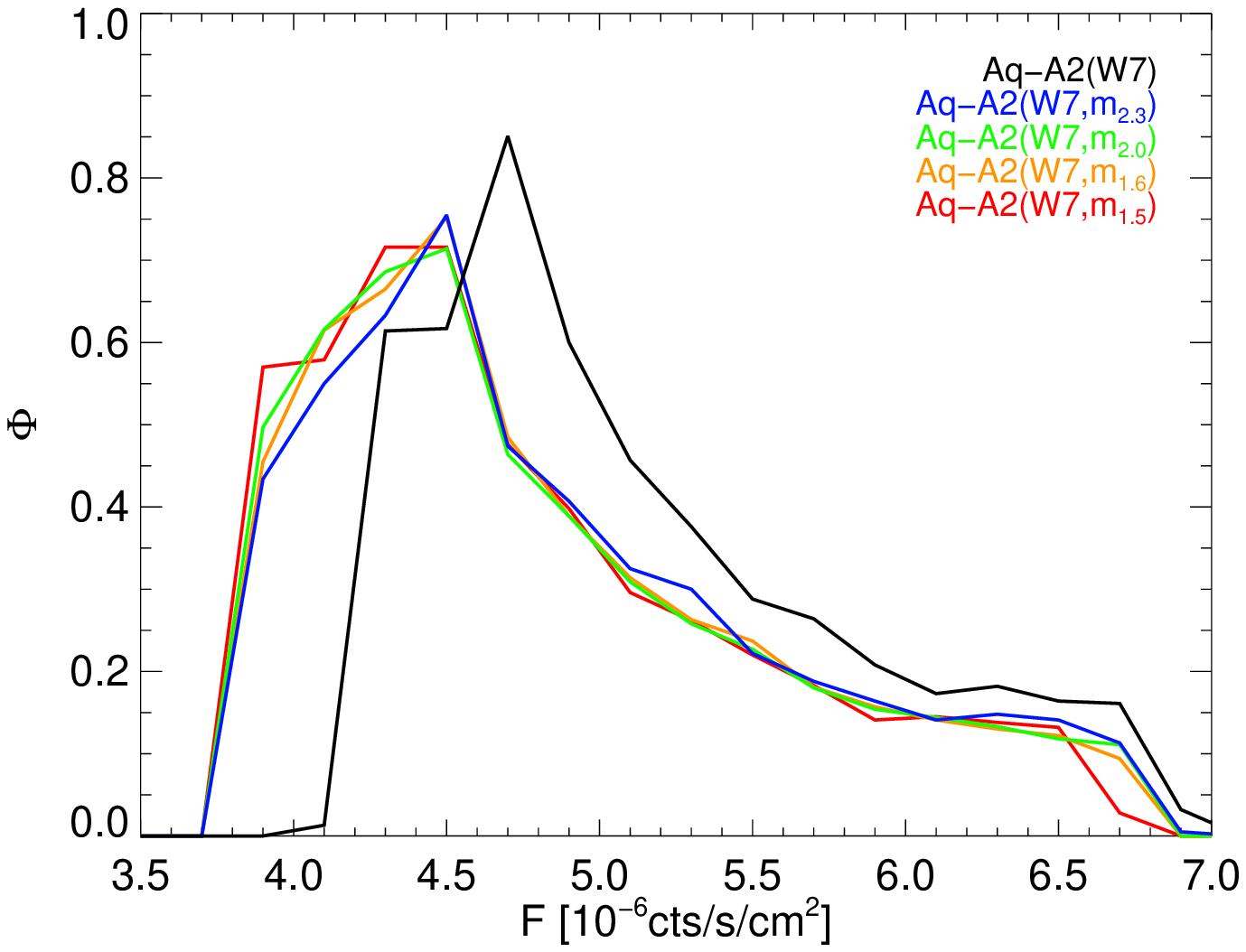}
   \includegraphics[scale=0.35,angle=-90]{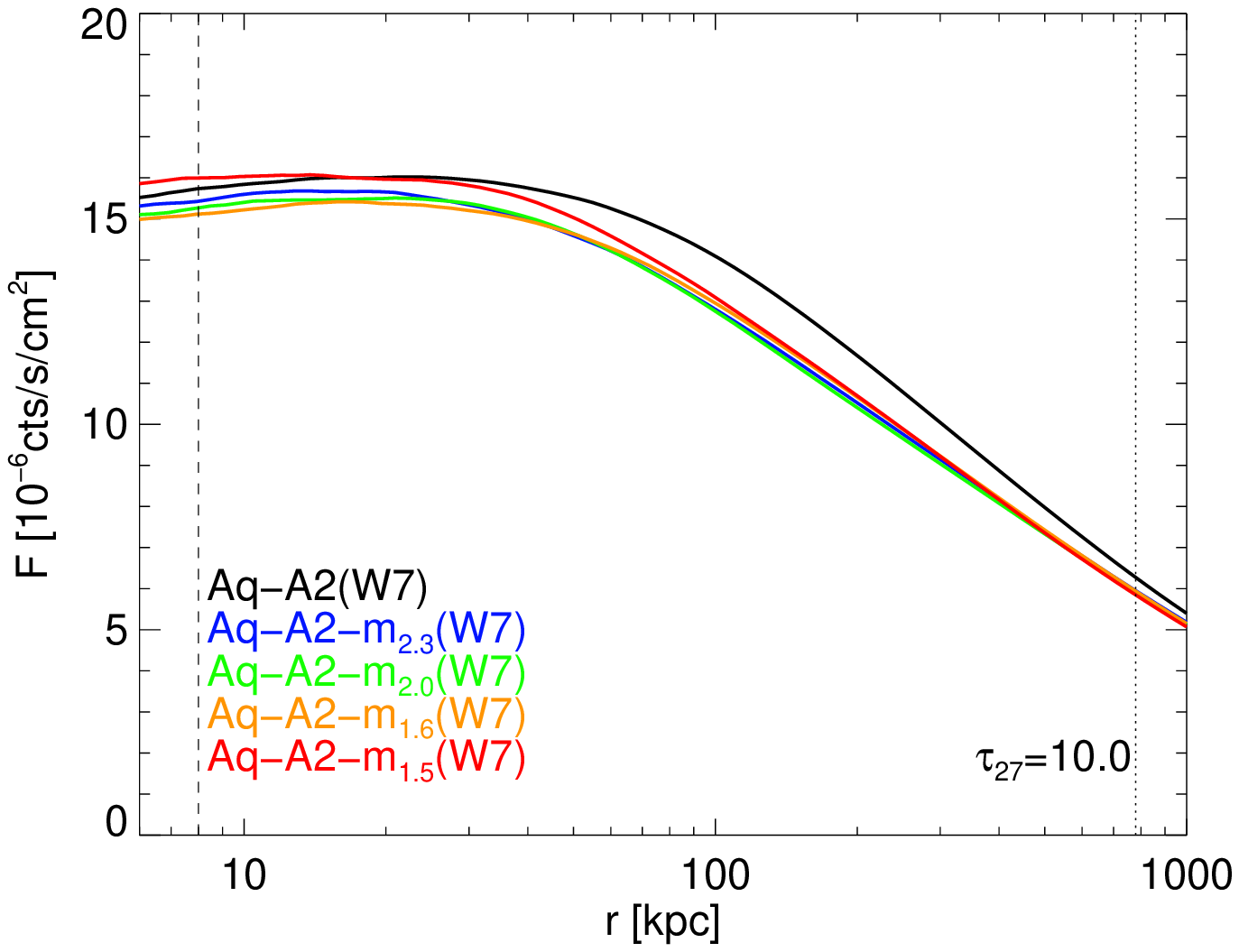}
   \caption{Impact of WDM. \emph{Left:} The M31 flux distributions for Aq-A2(W7) as
     simulated with CDM (black) and WDM models of thermal relic particle
     masses 1.5kev (red), 1.6keV (orange), 2.0keV (green) and
     2.3keV(blue). \emph{Right:} The flux surface density as a function
     of radius for these five models.}
   \label{FigSLWDM}
 \end{figure*}

 We now address the effect of changing the primordial matter power spectrum. One
 well-motivated candidate for decaying dark matter is a
 resonantly-produced sterile neutrino \citep{Shi99,Laine08,Boyarsky09a}. A sterile neutrino with a
 mass of 7~keV has a non-negligible free-streaming length that erases small
 scale power in the early Universe. The resulting matter power
 spectrum therefore possesses a cutoff similar to that of warm dark
 matter (WDM). The position and slope of the cutoff is not uniquely
 determined by the sterile neutrino mass, as the sterile neutrino
 momentum distribution is modified in the presence of a lepton
 asymmetry, which is a relatively unconstrained parameter. The family of
 spectra for a sterile neutrino mass of 7~keV and an unconstrained
 lepton asymmetry has cutoffs in the range between those of a 2.0~keV and a
 3.3~keV thermal WDM candidate
 \citep[][Lovell~et~al. in prep.]{Abazajian14}. WDM models of this
 type have a considerable effect on the
 structure of dwarf galaxies
 \citep{Lovell12,Maccio12,Maccio13,Shao13,Schneider14}, and perhaps to
 a much smaller extent, on that of Milky Way-analogue haloes. In
 Figure~\ref{FigSLWDM} we plot the M31 sightlines measurement and the
 surface density flux as a function of radius for our {\it WMAP7} version of Aq-A2 (CDM) and four WDM models (see
 figure caption).    

 The WDM flux distributions are suppressed relative to CDM for
 the M31 measurement by the order of a few per cent. The variation
 between WDM models is by comparison very small. The 1.5keV model is
 further suppressed than the 2.3keV, suggesting that flux correlates
 inversely with the free-streaming length, albeit weakly for this
 range of models. The discrepancy between CDM and the WDM
 models is reflected in the flux surface density profile, where CDM has
 a notably higher flux amplitude for observer distances greater than
 40kpc. At the distance of M31 this again amounts to a few
 per cent. At the GC observer distance the separation is much smaller
 and no longer correlates consistently with dark matter free-streaming length. This
 result is reflected in the sightlines distribution for the GC (not shown). The
 7keV sterile neutrinos will likely have a cooler power spectrum
 than that of even the 2.3keV thermal relic, therefore we do not
 consider further the effect of the dark matter free-streaming
 length on the flux distribution.

 \subsection{Baryonic physics}
\begin{figure}
   \includegraphics[scale=0.35,angle=-90]{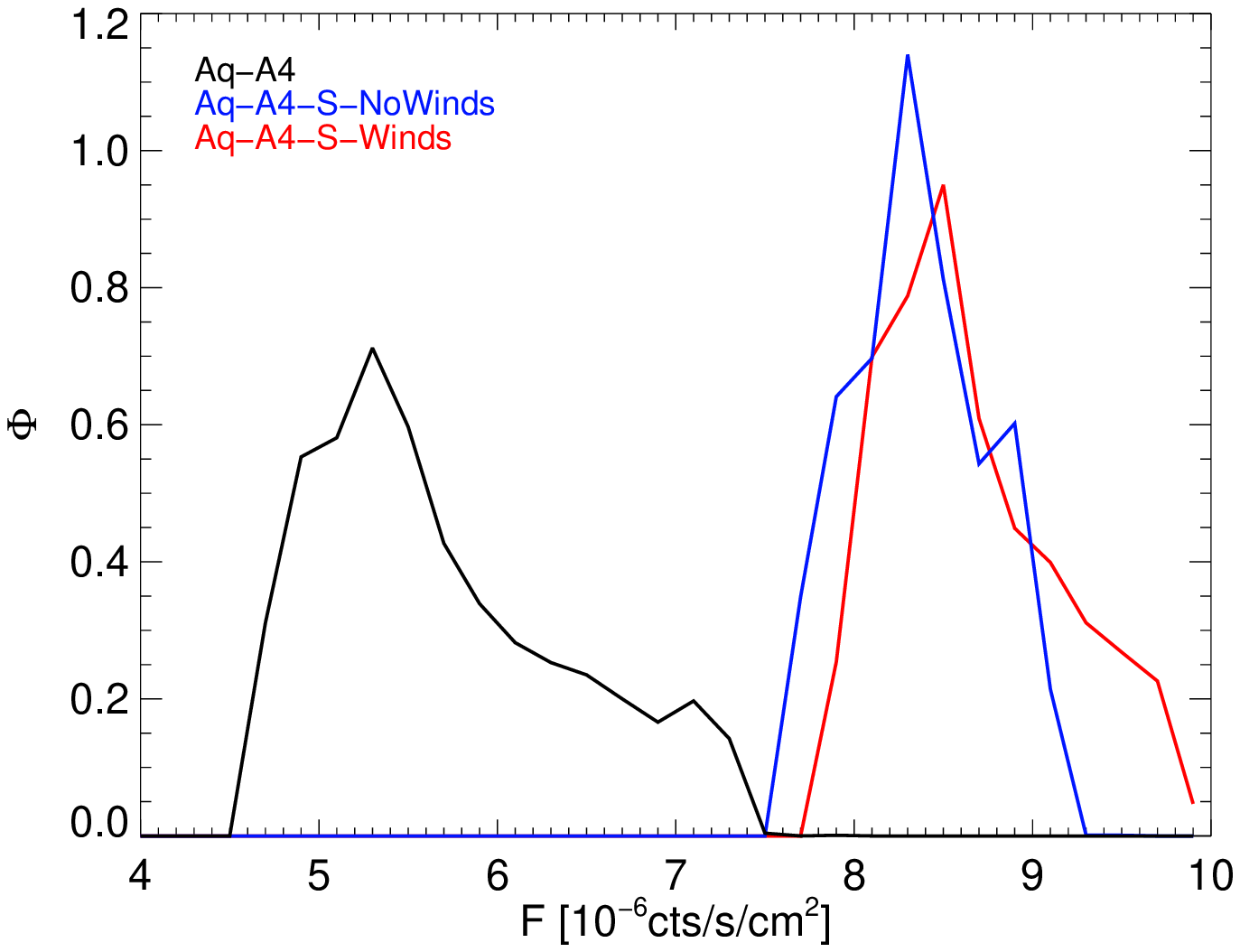}
    
   \caption{Impact of baryons. \emph{Left:} The flux distributions from M31 for Aq-A4 as
     simulated with just dark matter (black), and two gas-hydro models:
     red includes winds and blue does not}. 
   \label{FigSLBa}
 \end{figure}

 One crucial component of the cosmological model that is missing from
 the simulations presented so far is baryonic physics. Through the action of
 feedback and adiabatic
 contraction it is possible for baryons to alter considerably
 the distribution of dark matter and hence the expected X-ray flux
 \citep{Blumenthal86}. We
 have run realisations of Aq-A4 ({\it WMAP1}) with two baryonic models. Both
 make use of the star-formation and gas physics prescriptions
 introduced in \citet{Springel03}; one includes galactic winds and the other
 does not. With these simulations we are also able to relax the
 assumption that DM and baryons have the same spatial distribution. The M31 flux distribution functions for these simulations along with
 that of dark matter-only Aq-A4 are plotted in Figure~\ref{FigSLBa}.

 The variation in flux distributions is substantial. Both baryon model
 simulations
 exhibit M31-distance flux distributions that are $\sim50$ per cent
 higher than that of the DM-only run. Alternative gas/star-formation physics
 and subgrid recipes may
 produce different results \citep[e.g.][]{Schaller14, Mollitor15}: here we simply state that the precise
 effect of gas physics is uncertain and likely to effect the outcome of
 our measurement for M31. Due to the poor resolution of the GC
 measurement in Aq-A4 we do not attempt to draw conclusions about the
 likely effect of baryons on the GC results. For the same reason,
 we also do not attempt to use the baryonic simulations for the dwarf
 spheroidals; the precise effect of
 baryonic physics on dwarf galaxy mass distribution has been highly
 debated in the literature \citep[c.f.][]{diCintio12,GarrisonKimmel13,Brooks2014}.

 \subsection{Halo sample variance}
\begin{figure*}
   \includegraphics[scale=0.35,angle=-90]{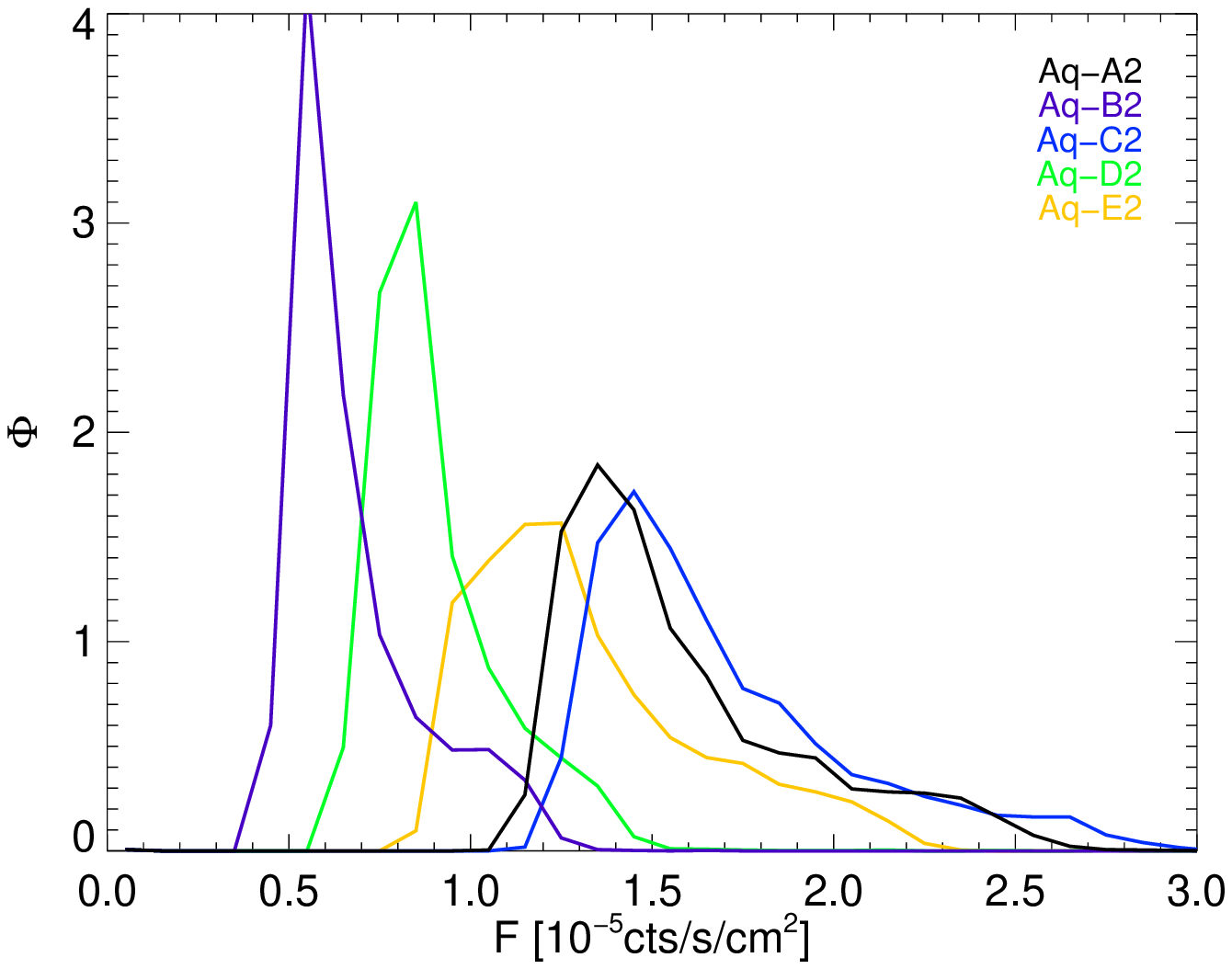}
   \includegraphics[scale=0.35,angle=-90]{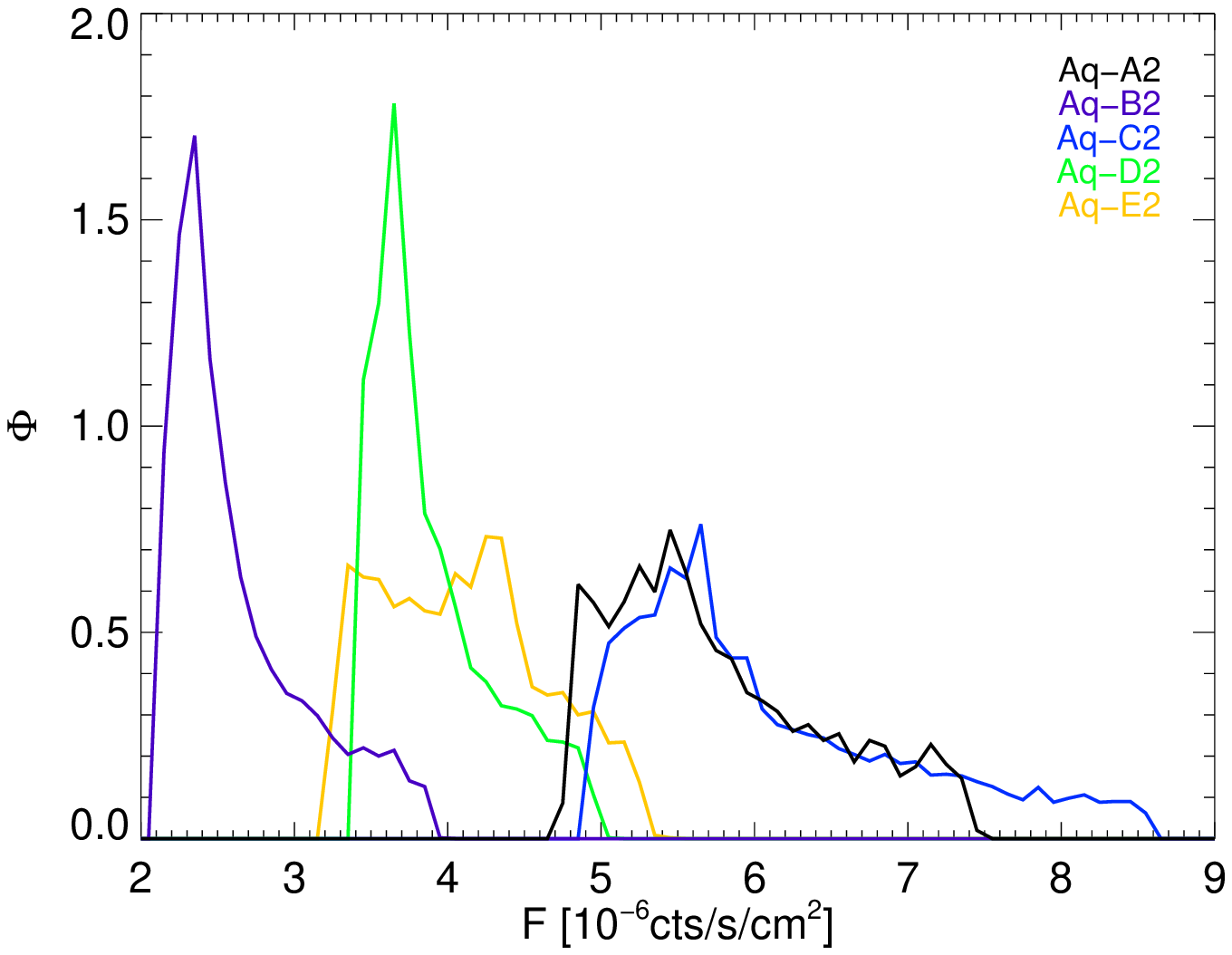}
   
   \caption{The flux distributions for Aq-A2 (black), Aq-B2
     (purple), Aq-C2, (blue), Aq-D2 (green) and Aq-E2 (orange). The left
     panel is for the GC, the right for M31.}
   \label{FigSLABCDE}
 \end{figure*}

 The measured flux will be very sensitive to the concentration of both
 the MW and M31 haloes, which decreases with halo mass albeit with
 considerable scatter \citep{Gao_04, Neto07}. Also, individual haloes will have very stochastic
 formation histories, and so will likely have very different
 structures. We therefore use four further Aquarius haloes (B-E) to
 examine the scatter in flux measurements that is likely to result
 from these variations in halo properties.  In Figure~\ref{FigSLABCDE}
 we plot the GC and M31 flux distributions for Aq-A2 plus these extra
 four haloes, also simulated at resolution level 2.

 The GC flux distribution of the Aq-C2 halo is enhanced by over a
 factor of two relative to Aq-B2 (the least massive halo). It is also a factor of
 two larger than is the case for Aq-D2, which is remarkable in that the
 Aq-D2 and Aq-C2 haloes have the same mass to at least three
 significant figures. The underlying difference is halo concentration:
 the value of the
 concentration parameter $c$ is 18.4 for Aq-C2 and 12.4 for
 Aq-D2. The difference between haloes is even larger
 for the M31 case: therefore, the variation in halo concentration
 between our haloes will be very important for our results.  

 \subsection{Position of observer}
 \label{PoO}
\begin{figure}
   \includegraphics[scale=0.35,angle=-90]{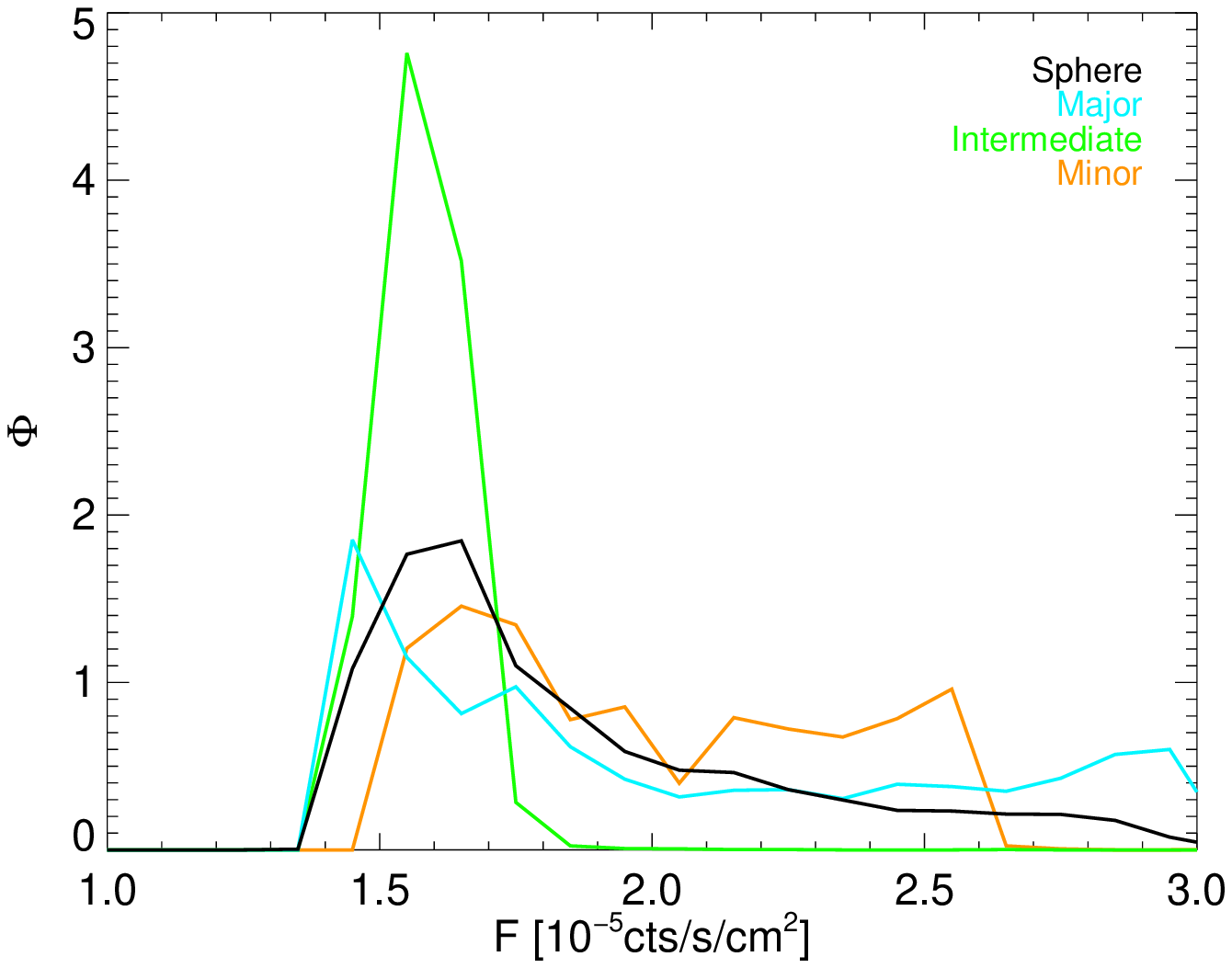}  
   \caption{Flux distributions for the GC (Aq-A1) when the observer
     is constrained to lie in the plane normal to the minor (orange),
     intermediate (green) and major (cyan) axes; the spherically uniform
     sample is reproduced in black.}
   \label{FigSLAxes}
 \end{figure}

 The final effect that we consider is that of the position of the
 observer within the Milky Way halo. It has been shown that the
 stellar disc is at its most stable when aligned with the inner dark matter
 halo's minor axis \citep{Bailin05,Aumer13}, and that the orientation of
 these vectors is constant from the inner resolution limit to
 $0.1\times r_{200}$ \citep{VeraCiro11}. We therefore calculate the inertia tensor of all
 the dark matter particles within the sphere of radius $0.1r_{200}$,
 extract the minor axis vector for the corresponding ellipsoid, and
 then randomly distribute observers in rings of radius 8kpc with the minor axis as the normal
 vector. We plot the distribution functions in
 Figure~\ref{FigSLAxes}; for completeness we also include the results when the
 intermediate and major axes are used as the normal vector. The simulation
 used is Aq-A1.

 The different axes have a noticeable effect on the distribution of
 fluxes. The shapes of the constrained-observer curves have quite
 different shapes to the the spherical-sampling case. For the major
 axis measurement the distribution at low fluxes is remarkably similar
 to the spherical case, but is then skewed heavily towards higher
 fluxes. The minor axis has consistently higher fluxes than does
 spherical sampling, by up to a few tens of per cent. Given that the
 effect is non-negligible, we will therefore
 build the possiblity that our observer is biased towards the minor plane
 axis into our final results.

\section[]{Analysis of XMM-Newton/EPIC observations of Draco dwarf spheroidal galaxy}
\label{DO}

 In this section we describe the details of our analysis of the Draco dwarf
spheroidal galaxy (dSph) as observed with the European Photon Imaging
Camera (EPIC) on-board the {\it XMM-Newton} space mission. Based on these
observations, we fail to detect any significant line at $\sim
3.5$~keV, thus placing an upper bound on its strength. Finally, we
estimated the exposure of Draco observations by {\it XMM-Newton}
necessary to confirm the dark matter origin of the $\sim 3.5$~keV line.
The results of this section have also been used for {\it XMM-Newton}
proposal \#076480, which was accepted in {\it XMM-Newton} 
14th Announcement of Opportunity (AO-14); 
see \url{http://xmm.esac.esa.int/external/xmm_news/otac_results/ao14_results} for details.

\subsection[]{Data reduction}

 In this paper we use data from the publicly available {\it XMM-Newton}
observations of the Draco dSph (ObsIDs 
\texttt{0603190101}, \texttt{0603190201},  \texttt{0603190301}, \texttt{0603190401} and \texttt{0603190501}). 
Initial data files from the MOS and PN cameras of XMM-Newton/EPIC are 
pre-processed with the \texttt{emproc} and \texttt{epproc} procedures
of the standard {\it XMM-Newton} software 
SAS v.13.5.0. We detect data patterns with significant spatial (bright point sources) 
and temporal (proton flares) variabilities using the standard SAS procedures \texttt{espfilt} and 
\texttt{edetect\_chain}, remove these patterns from the subsequent analysis,
and extract source spectra from 14' radius circle centred on Draco dSph centre 
($\mathrm{RA}={}$17:20:12.4, $\mathrm{DEC}=+$57:54:55.3) using SAS procedure \texttt{evselect}.
Redistribution matrix files (RMF) and ancillary response files (ARF)
are created with the standard SAS 
procedures \texttt{rmfgen} and \texttt{arfgen}, respectively. Finally, we group the obtained 
source spectra and response files, co-add them channel-by-channel using FTOOLS procedure 
\texttt{addspec}, and rebin the obtained spectra by 60~eV ($\sim 2-3$ times smaller than the instrument's energy 
resolution) to make the energy bins roughly statistically independent. 
The total cleaned exposure of the obtained spectra is 107.1~ks for MOS (either MOS1 or MOS2) 
cameras and 40.4~ks for PN.

\subsection[]{Spectral modelling}

 We model the obtained Draco spectra in the 0.8-10.0~keV range
  using the X-ray spectral fitting package \texttt{Xspec} 
v.12.8.1g. Because previous studies of dwarf spheroidal galaxies have
  not revealed the presence of any X-ray emitting gas, our model is a
  sum of instrumental and astrophysical background components.  No
  signature of residual soft protons has been found according to the
  procedure of~\citet{2004A&A...419..837D}, so we have not added the
  residual soft proton component. The instrumental background (mostly caused by cosmic MeV protons penetrating inside {\it XMM-Newton satellite}) is modelled by an unfolded 
\texttt{powerlaw} model and the sum of several narrow \texttt{gaussian}s representing bright fluorescence 
lines. The astrophysical background was modelled by a sum of the cosmic X-ray background (a folded 
\texttt{powerlaw}) and the Galactic X-ray background (two \texttt{apec} models), in full accordance 
with~\citet{Malyshev14}. The resulting fit quality is good, with $\chi^2$ = 221.64 for 223 d.o.f.
Then, by adding further narrow \texttt{gaussian} lines, we looked for line-like 
residuals in our region of interest near 3.5~keV. No statistically significant residuals were found.
This produces a 2$\sigma$ upper bound of $5.6\times 10^{-6}$
  cts/s/cm$^2$ on extra line flux in the 3.45-3.58~keV range. 

 \begin{figure*}
     \includegraphics[width=0.3\textwidth,angle=-90]{Figures/draco_spectrum}
   \includegraphics[width=0.3\textwidth,angle=-90]{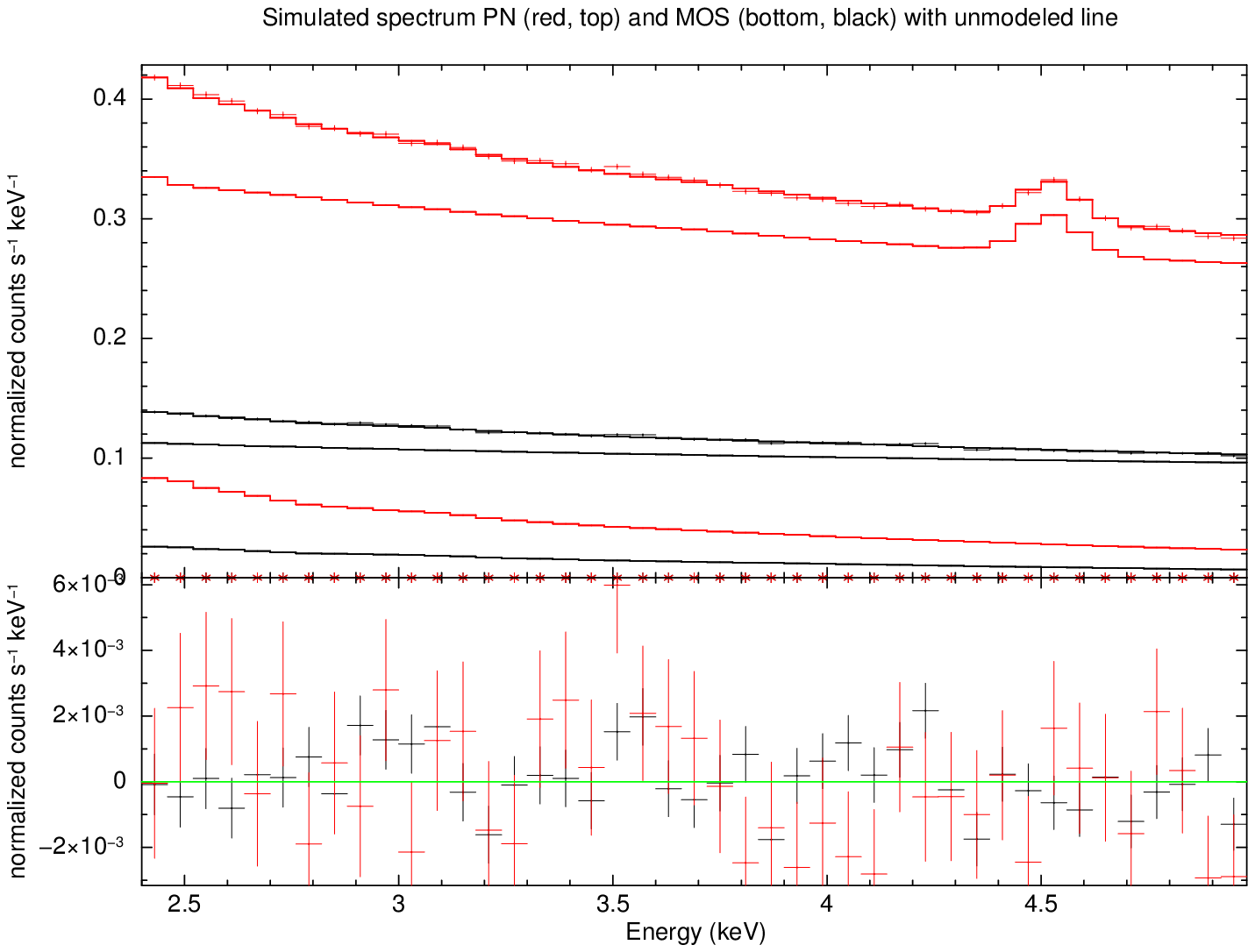}
   \caption{\textit{Left}: 
  Combined spectrum of existing Draco dSph observations
   modelled as a combination of folded components (absorbed thermal low energy
    Galactic emission), an extragalactic \texttt{powerlaw} (sharply falling
    component), and an instrumental component (unfolded \texttt{powerlaw} plus
    instrumental Gaussians). The quality of fit: $\chi^2 = 221.64$ per 223
    d.o.f. \textit{Right}: simulated spectrum of $1.34$~Ms of Draco
    dSph. The line with the flux $F_\text{Draco,min}$ is detected in two
    cameras with combined $\Delta \chi^2 = 13.0$.}

 \label{fig:draco}
 \end{figure*}

\subsection[]{Calculation of sensitivity with respect to narrow line at $\sim$3.5~keV}

To determine the exposure of Draco observation necessary to check the decaying dark matter 
hypothesis of the $\sim$3.5~keV line, we first calculate the expected line strength. 
According to Figure~\ref{DraScu}, 
the best-fit ratio $F_\text{Draco}/F_\text{GC} = 0.09$ with the scatter ranging from 0.04 to 0.2
(95 per cent range). Taking the results of~\citet{Boyarsky14b}, where the line was
detected from the GC with the highest significance, the best-fit line flux of
$\unit[26 \times 10^{-6}]{cts/s/cm^2}$ leads us to the conclusion that
the range of fluxes expected from the Draco dSph is $ \unit[(1.0-5.2)\times
10^{-6}]{cts/s/cm^2}$ with the most plausible value being $
\unit[2.3\times 10^{-6}]{cts/s/cm^2}$. We take the value
$F_\text{Draco,min}=1.0\times 10^{-6} \unit{cts/s/cm^2}$ as a conservative lower bound on 
the expected DM signal in Draco.

The existing observations of Draco (Fig.~\ref{fig:draco}, left panel) allow us
to determine the count rate at the energies of interest. It shows that one
expects $N_\text{bg} = 5.79\times 10^4$~cts (two MOS cameras combined) or
$N_\text{bg} = 8.36\times 10^4$~cts (PN camera) from a 1.34 Ms
observation in a 180 eV energy interval (corresponding to the broadening of a
narrow line due to the spectral resolution of XMM). Using the same exposure
and the expected DM line flux of $1.02\times 10^{-6}\unit{cts/s/cm^2}$ we
find $N_\dm = 528$~cts for the MOS cameras (combined MOS1 and MOS2) and $N_\dm
= 600$~cts for the PN camera. Therefore the expected significance of the
signal against this background is $N_\dm/\sqrt{N_\text{bg}}=
(528+600)/\sqrt{(5.79+8.36)\times 10^4} \approx 3.0$.

To make this conclusion more robust we performed simulations of
long-exposure observations. First, using the \texttt{fakeit} command of \texttt{Xspec}, we simulated realizations of the Draco dSph spectrum with the line added at
  the $\unit[10^{-6}]{cts/s/cm^2}$ level and with a {\it XMM-Newton} exposure 1.34~Ms: we recovered the line
at a more than $3\sigma$ level. An example of the simulated spectrum with a
positive line-like residual at $\sim 3.5$~keV is show in Fig.~\ref{fig:draco},
right panel. 


We also generate 250 realizations of spectra of a $1$~Ms observation of the
Draco dSph (based on the model of the existing Draco data, see
Fig.~\ref{fig:draco}). The simulated spectra do not contain a line at
  $E\approx 3.5$~keV. We then try to detect a line at this energy and find
that only in 12 simulations (i.e.\ in $4.8$ per cent of cases) were we able to detect
a line-like residual at a level of $\unit[10^{-6}]{cts/s/cm^2}$ or
above. This supports our estimate that a $1$~Ms observation will
either confirm the existence of the line or will instead rule it out at least at the
$95$ per cent confidence level.

In addition we have simulated 100 realizations of the Draco dSph spectrum
with the line added at the $\unit[10^{-6}]{cts/s/cm^2}$ level. In
68 per cent of the realizations the line was recovered with a flux in the range $\unit[(0.71
-1.45)\times 10^{-6}]{cts/s/cm^2}$, consistent with the Gaussian scatter
around the simulated value.

\end{document}